\newcommand{\version}{April 16, 2018}
\documentclass[letterpaper,oneside,11pt,pdftex]{article}
\pdfoutput=1
\usepackage[bindingoffset=0.0cm,textheight=22.5cm,hdivide={*,16cm,*}, vdivide={*,22.5cm,*}]{geometry}
\usepackage{amsmath,amsfonts,amssymb}
\usepackage{mathtools}
\usepackage[pdftex]{graphicx}
\usepackage[margin=1cm,font=small]{caption}
\usepackage[T1]{fontenc}
\usepackage[utf8]{inputenc}
\usepackage{fouriernc}
\newcommand{\id}{\mathbb{1}}
 \usepackage[numbers,square,comma,sort&compress]{natbib}
\usepackage[pdftex,hyperref,svgnames]{xcolor}
\usepackage[pdftex,bookmarksnumbered=true,breaklinks=true,%
colorlinks=true,linktocpage=true,linkcolor=MediumBlue,citecolor=ForestGreen,urlcolor=DarkRed]{hyperref}

\makeatletter
\AtBeginDocument{
  \hypersetup{
    pdfkeywords = {\keywords},
    pdftitle = {\@title},
    pdfauthor = {\@author}
  }
}
\makeatother

\makeatletter

 \@addtoreset{equation}{section}
 \makeatother

\renewcommand{\b}{\beta}
\newcommand{\g}{\gamma}
\renewcommand{\d}{\delta}
\newcommand{\e}{\epsilon}

\newcommand{\vth}{\vartheta}

\renewcommand{\l}{\lambda}
\newcommand{\m}{\mu}
\newcommand{\n}{\nu}

\newcommand{\s}{\sigma}

\newcommand{\ph}{\phi}

  \newcommand{\cO}{\mathcal{O}}

\newcommand{\pa}{\partial}

\newcommand{\nn}{\nonumber}
\newcommand{\eqnref}[1]{eq. \eqref{#1}}

\newcommand{\diff}[2]{\frac{\pa #1}{\pa #2}}

\newcommand{\rd}{\mathrm{d}}
\newcommand{\txt}[1]{\textrm{#1}}

\newcommand{\ct}{c_{\textrm{t}}}
\newcommand{\cl}{c_{\textrm{l}}}
\newcommand{\cRl}{c_{\textrm{R}}}
\newcommand{\gt}{\gamma_{\textrm{t}}}
\newcommand{\gl}{\gamma_{\textrm{l}}}
\newcommand{\bt}{\beta_{\textrm{t}}}
\newcommand{\bl}{\beta_{\textrm{l}}}

\newcommand{\bm}{\beta_{\bar\mu}}

\newcommand*{\mat}[1]{\mathbf{#1}}
\renewcommand*{\mat}[1]{\mathrm{#1}}
\renewcommand*{\vec}[1]{\mathbf{#1}}
\newcommand{\lt}{\bar{\Gamma}}
\newcommand{\LT}{\Gamma}

\newcommand{\ftntmark}[1]{\texorpdfstring{\footnotemark[#1]~}{}}

\DeclarePairedDelimiter\abs{\lvert}{\rvert}
\newcommand{\coleq}{\vcentcolon=}
\title{\texorpdfstring{\begin{flushright}
        {\small LA-UR-17-29936%\\[-2ex]
	}
       \end{flushright}\vspace{2em}}{}%
Line tension of a dislocation moving through \texorpdfstring{\\}{}an anisotropic crystal}

\author{Daniel N. Blaschke\ftntmark{1} and Benjamin A. Szajewski\ftntmark{2}}

\date{\version}

\newcommand{\keywords}{moving dislocations, line tension, dynamic loading, anisotropy}

\graphicspath{{./}{./figures/}}

\begin{document}

\maketitle

\thispagestyle{empty}
\begin{center}
\renewcommand{\thefootnote}{\fnsymbol{footnote}}
\vspace{-0.3cm}
\footnotemark[1]Los Alamos National Laboratory, Los Alamos, NM, 87545, USA\\
\footnotemark[2]Army Research Laboratory, Aberdeen Proving Ground, MD, 21005, USA
\\[0.5cm]
\ttfamily{E-mail: dblaschke@lanl.gov, benjamin\_szajewski@alumni.brown.edu}
\end{center}

\vspace{1.5em}

\begin{abstract}
Plastic deformation, at all strain rates, is accommodated by the collective motion of crystalline defects known as dislocations.
Here, we extend an analysis for the energetic stability of a straight dislocation, the so-called \emph{line tension} ($\Gamma$), to steady-state moving dislocations within elastically anisotropic media.

Upon simplification to isotropy, our model reduces to an explicit analytical form yielding insight into the behavior of $\Gamma$ with increasing velocity.
We find that at the first shear wave speed within an isotropic solid, the screw dislocation line tension diverges positively indicating infinite stability.
The edge dislocation line tension, on the other hand, changes sign at approximately $80\%$ of the first shear wave speed, and subsequently diverges negatively indicating that the straight configuration is energetically unstable.

In anisotropic crystals, the dependence of $\Gamma$ on the dislocation velocity is significantly more complex;
At velocities approaching the first shear wave speed within the plane of the crystal defined by the dislocation line, $\Gamma$ tends to diverge, with the sign of the divergence strongly dependent on both the elastic properties of the crystal, and the orientation of the dislocation line.
We interpret our analyses within the context of recent molecular dynamics simulations (MD) of the motion of dislocations near the first shear wave speed.
Both the simulations and our analyses are indicative of instabilities of nominally edge dislocations within fcc crystals approaching the first shear wave speed.
We apply our analyses towards predicting the behavior of dislocations within bcc crystals in the vicinity of the first shear wave speed.
\\[2em]
\end{abstract}

\newpage
\tableofcontents
% \newpage

\section{Motivation}
\label{sec:intro}
%%%%%%%%%%%%%%%%%%%%%%%%%%%%%%%%%%%%%%%%%

Metallic plasticity is accommodated by the movement of linear crystalline defects known as dislocations.
In regimes of high-rate loading~\cite{Clifton:1983}, such as impact~\cite{Clifton:2000,Murr:1996}, the velocity of dislocations may approach the intrinsic elastic transverse wave speed of the continuum to accommodate high rates of material deformation.
A suite of recent molecular dynamics (MD) simulations~\cite{Olmsted:2005,Daphalapurkar:2014,Marian:2006,Oren:2017,Wang:2008,Gilbert:2011,Queyreau:2011} as well as some experimental data~\cite{Nosenko:2007} suggest that
dislocation velocities can reach or even exceed the transverse wave speed ($\ct$) of the material.
Within these \emph{dynamic} regimes, both the energetics and the stability of dislocations markedly differs from that as predicted within regimes of quasi-static deformation.
As an example, upon considering the interaction energy between a solid and a dynamic dislocation moving at velocities ($v$) approaching the transverse wave speed, discrete dislocation
theory predicts that the energy of nominally screw and edge dislocations diverges as $\sim \mathcal{O}(1-(v/\ct)^2)^{-1/2}$ and $\sim \mathcal{O}(1-(v/\ct)^2)^{-3/2}$, respectively~\cite{Weertman:1961,Weertman:1980}.
Both experiment and numerical simulations suggest that dislocations may traverse at, or even exceed, the transverse wave speed of the continuum.
On the other hand, consideration of dislocation theory suggests divergent energies in these regimes.
Therefore, the notion of straight, steady state dislocation motion at velocities approaching the transverse wave speed requires reconciliation.

Central to stability analyses of dislocations is the dislocation line tension ($\LT$).
Analogous to a taut string, dislocations exhibit a tendency to straighten, minimizing their length, and also their interaction energy with the crystal.
% The line tension, as defined in \eqnref{eq:def-linetension} below,
The line tension
is a scalar quantity related to the configurational force acting to straighten and minimize the length of a dislocation, if the line tension is positive (indicating stability of the configuration)~\cite[sec. 2.6]{Bacon:1980}.
Conversely, a negative line tension indicates instability, and the related force will drive an unstable dislocation away from its current configuration.
The line tension depends upon the orientation, i.e. the angle ($\vth$) between the dislocation Burgers vector ($\vec{b}$) and the dislocation line direction ($\vec{t}$)~\cite{Hirth:1982,Bacon:1980}.
Including an orientational dependence in the small bowout limit, as with anisotropic surfaces~\cite{kingston1951physics}, or dislocations as considered here, the line tension ($\Gamma$) is related to the (dislocation) energy $E(\vth,v)$ as~\cite{DeWit:1959}
\begin{align}
 \LT&=\left(1+\frac{\pa^2 }{\pa \vth^2}\right)E(\vth,v)
 \,, \label{eq:def-linetension}
\end{align}
where $v$ denotes the dislocation velocity.
Eq.~\eqref{eq:def-linetension} is valid for curved as well as for straight dislocations (which are merely a special case of the former)~\cite[sec. 2.6]{Bacon:1980}.
Within this work, we assume a bow-out type mechanism of instability which occurs along a dislocation glide plane.
Independent of the details of the mechanism, for a characteristic dislocation length, $\mathcal{L}$, the dislocation line tension \eqref{eq:def-linetension} reduces to~\cite[pp.~174--177]{Hirth:1982}
\begin{align}
 \LT&=\lt\frac{\bar{\mu}b^2}{4\pi}\ln \left(\frac{\mathcal{L}}{r_0}\right)+\Gamma_0
 \,, \label{eq:def-bowout}
\end{align}
where $\lt\coleq\lt(\vth,v)$ denotes the dislocation line tension prelogarithmic factor,
% $\bar{\mu}=(c_{11}-c_{12}+2c_{44})/4$
$\bar\mu$ is the average shear modulus,
and $r_0$ represents the cutoff parameter introduced to circumvent divergent dislocation energies.
In addition, the term $\Gamma_0$ depends upon the geometry of the bow-out under consideration (e.g., small bow-out~\cite{Hirth:1982}).
The two constants $r_0$ and $\Gamma_0$ depend upon details of the dislocation core
% which are beyond the scope of discrete linear elasticity,
and details of the specific bow-out mechanism under consideration, respectively,
both of which are beyond the scope of the present work.
In the limit of an infinitely long dislocation ($\mathcal{L}\gg r_0$), $\Gamma\sim(\bar{\mu} b^2/4\pi)\lt\ln(R/r_0)$, with $R$ (of the order of interdislocation spacing) denoting the radius of an outer cylinder surrounding the dislocation core, however the magnitude of $\lt$ does not depend upon $\mathcal{L}$, $R$, or detailed geometrical features of the bow-out. Within this work we focus our attention on $\lt$ \textit{only} within \eqnref{eq:def-bowout}, since it represents the contribution of the cumulative strain and kinetic energy of the crystal to the dislocation line tension, and at $\mathcal{L}\gg r_0$ is expected to dominate within \eqnref{eq:def-bowout}.

The objective of this work is to study the dependence of the dislocation line tension prelogarithmic factor $\lt$ on both the velocity and orientation of the dislocation line (i.e. $\lt(\vth,v)$).
The dependence of the dislocation line tension on the velocity provides a mechanistic interpretation of the critical velocity at which a moving (straight) dislocation becomes energetically unstable with respect to bow-out.

As noted, MD simulations for fcc metals such as~\cite{Marian:2006,Daphalapurkar:2014,Oren:2017} show
\footnote{
One exception is ref.~\cite{Wang:2008}, as the authors discussed only screw dislocations and naturally no jump was found there.
}
that at some critical driving stress the dislocation velocity ``jumps'' to the transonic regime, exhibiting a discontinuity in the velocity response to an applied stress, (and a similar jump to supersonic happens around the longitudinal sound speed).
In fcc metals these rapid shifts in velocity are limited to edge dislocations \emph{only}, where the Burgers vector is perpendicular to the dislocation line, indicating that the character of the dislocation is important towards understanding the mechanics of dislocation motion within high velocity regimes.
Furthermore, there remains sustained interest~\cite{Blaschke:BpaperRpt} in delineating the limits of applicability of the dislocation drag coefficient to infinite, straight, dislocations in regimes of high dislocation velocity.
Current models~\cite{Alshits:1992,Blaschke:BpaperRpt} assume perfectly straight dislocations and thereby introduce artificial (unphysical) divergences at the transverse sound speed within an infinite isotropic crystal.

Some authors~\cite{Markenscoff:2009} have demonstrated, that if terms accounting for acceleration are included into the models of screw dislocation fields in the isotropic limit, the divergence that is otherwise present at the transverse sound speed in the steady state approximation, disappears.
While this may explain the smooth transition of screw dislocations into the transonic regime, it does not explain the sudden jump-like acceleration from sub- to transonic observed for edge dislocations at some critical stress~\cite{Marian:2006,Daphalapurkar:2014,Oren:2017}.
% It is this latter effect that we aim at explaining here.

%%% Our Mechanistic Interpretation.
Our plausible interpretation of these findings is as follows.
At some critical velocity which is a fraction of the transverse sound speed, the (edge) dislocation line tension (eq. \eqref{eq:def-linetension}) becomes negative, favoring bow-out of the dislocation line.
With the configurational change of the dislocation line, the drag coefficient which relates driving stress to dislocation velocity also decreases;
The dislocation, which is no longer straight, then experiences an abrupt acceleration into the transonic regime.
This description is qualitatively in agreement with recent MD simulations, e.g.~\cite{Oren:2017}, where a bowing nucleus is observed to form on a transient dislocation at the instant of acceleration.
Hence, the objective presently is to generalize the work on anisotropic line tension of ref.~\cite{Barnett:1972} to include a velocity dependence along the same lines as was previously done only within the isotropic case~\cite{Sakamoto:1991} (where edge and screw dislocations decouple making the orientation dependence of $\lt$ trivial).
This will provide an initial estimate for the evolution of the line tension with increasing velocity, leading to a critical velocity, beyond which a straight dislocation is energetically unstable
\footnote{
What that new shape is exactly will be more difficult to determine and may be pursued in future studies.
}.
Our present study will be limited to the line tension calculated from the dislocation self-energy, i.e. we consider a single infinitely long dislocation and infinitesimal bowout, leaving the contribution from dislocation-dislocation (and dislocation-crystal lattice) interactions to future work.

The remainder of this work is organized as follows.
Within the next section, we begin by reviewing the equations governing the dynamic balance of linear momentum for a linear elastic solid.
We then provide a brief review~\cite{Stroh:1962} of the elastic field equations for an infinite dislocation situated within a linearly elastic anisotropic crystal of either fcc or bcc structure.
With the elastic field equations, and the governing equations of dynamic equilibrium we are able to formulate expressions for both the energy and line tension of a dislocation within an anisotropic crystal with respect to both the orientation ($\vth$) and the velocity ($\vec{v}$) of the straight dislocation.
Within section~\ref{sec:isotropic}, we consider the limit of elastic isotropy, demonstrating that our expressions for both the energy and line tension reduce to already existent expressions available within the literature~\cite{Sakamoto:1991}.
Within section~\ref{sec:cubic}, we apply our expression for the line tension to several anisotropic fcc and bcc 
structured crystals examining the dependence of $\lt$ on both the velocity and orientation of the dislocation line.
In the final section~\ref{sec:conclusion}, we provide a brief synopsis of the results of this work.

\section{Method}
\label{sec:method}
%%%%%%%%%%%%%%%%%%%%%%%

We proceed with a brief summary of the general method for deriving displacement gradient fields of dislocations.
Upon integrating over the displacement gradient fields, both the dislocation energy and line tension may be obtained.
More details can be found within the extensive reviews~\cite{Hirth:1982,Bacon:1980}.

\begin{table}[h!t!b]
\centering
\caption{\label{tab:values-metals}We list experimentally determined values of both elastic constants and densities of various metals at room temperature used within our computations.
 All values may be found within refs.~\cite{Haynes:2017,Thomas:1968,Epstein:1965,Leese:1968,Bolef:1961,Soga:1966,Lowrie:1967}.
 %refs. 1, 15 therein
 The last column lists the Zener anisotropy ratios $A\coleq 2c_{44}/(c_{11}-c_{12})$ computed from the listed elastic constants.}
\begin{tabular}{lccccc}
\hline
\hline
\vspace{0.05cm}
	&	$\rho$[g/ccm]	&	$c_{11}$[GPa]	&	$c_{12}$[GPa]	& $c_{44}$[GPa]	&	A\\
\hline 
fcc	&			&			&		&	&	\\
\hline
Al	&	2.70	&	106.75	&	60.41	&	28.34	&	1.22 \\
Cu	&	8.96	&	168.3	&	122.1	&	75.7	&	3.28 \\
Ni	&	8.90	&	248.1	&	154.9	&	124.2	&	2.67 \\
\hline
bcc	&			&			&		&	&	\\
\hline
Fe	&	7.87	&	226.0	&	140.0	&	116.0	&	2.70 \\
Nb	&	8.57	&	246.5	&	134.5	&	28.73	&	0.51 \\
Ta	&	16.4	&	260.2	&	154.4	&	82.55	&	1.56 \\
W	&	19.3	&	522.39	&	204.37	&	160.58	&	1.01 \\
\hline
\hline
\end{tabular}
\end{table}

Within the scope of this work, we are concerned with cubic (fcc and bcc) crystals, whose stiffness tensor contains the set of three cubic elastic constants $\{c_{11}, c_{12}, c_{44}\}$.
Tensors of elastic constants are typically defined in Cartesian coordinates which are aligned with the crystal axes.
For orientations of the coordinate system which are not aligned with the crystal bases, it is necessary to rotate the stiffness tensor into the prescribed basis.
The tensor of second order elastic constants (i.e., the fourth order elastic stiffness tensor) within the crystal reference frame reads
\begin{align}
 C_{ijkl}&=c_{12}\d_{ij}\d_{kl}+c_{44}\left(\d_{ik}\d_{jl}+\d_{il}\d_{jk}\right)-\left(2c_{44}+c_{12}-c_{11}\right)\sum_{\alpha=1}^3\d_{i\alpha}\d_{j\alpha}\d_{k\alpha}\d_{l\alpha}
 \,,\label{eq:STIFF}
\end{align}
where the sum over repeated indices is implicit, and the last term which includes the explicit summation represents the fourth order identity tensor.
Within table~\ref{tab:values-metals} we list the material specific elastic constants utilized within this study, as well as the material densities and Zener anisotropy ratio.
In the isotropic limit $c_{11}\to c_{12}+2c_{44}$,
and the three cubic constants reduce to the two Lam{\'e} constants: $c_{12}\to\l$, $c_{44}\to\m$.
The governing equations to which the displacement gradient field provides a solution are the equations of motion (e.o.m.) and the (leading order) stress-strain relations known as Hooke's law:
\begin{align}
 \pa_i\s_{ij}&=\rho\ddot u_j\,, &
 \s_{ij}&=C_{ijkl}\e_{kl}=C_{ijkl}u_{k,l}\,, \nn\\
 \e_{kl}&\coleq\frac12\left(u_{k,l}+u_{l,k}\right)
 \,, \label{eq:Hooke}
\end{align}
where $\e_{ij}$ denotes the infinitesimal strain tensor, $\rho$ the material density, and the last equality in the first line follows from the elastic constants' Voigt symmetry.
Furthermore, we have introduced the notation
\footnote{
Since we are considering Cartesian coordinates we do not worry about co-/contra-variant indices.}
$u_{k,l}\coleq\pa_l u_k$ for the gradient of the displacement field $u_k$, and $\ddot u_j\coleq \frac{\pa^2 u_j}{\pa t^2}$ denoting temporal derivatives.
In the static case $\ddot{\vec{u}}=\vec{0}$, however even at constant, finite dislocation velocity a similar simplification may be devised.
In the dynamic case with constant velocity, the displacement field may be expressed in terms of a time-dependent shifted position $(\vec{x}-\vec{v}t)$, i.e. $u_k(x_i, t)=u_k(x_i-v_i t)$, and hence its temporal derivative can be expressed in terms of its gradient: $\dot u_i=-v_j u_{i,j}$.
Then the e.o.m. (eq. \ref{eq:Hooke}) may be simplified as
\begin{align}
 0&=\pa_i\s_{ij}-\rho\ddot u_j=C_{ijkl}u_{k,il}-\rho v_iv_l u_{j,il}
 \,,
\end{align}
or upon defining ``effective'' elastic constants (or the ``dynamic stiffness tensor'')~\cite{Bacon:1980}:
\begin{align}
 \hat C_{ijkl}u_{k,il}&=
%  \hat C_{ijkl}\pa_i u_{k,l}=
 0\,, &
 \hat C_{ijkl}\coleq C_{ijkl}-\rho v_iv_l \d_{jk}
 \,. \label{eq:eom1}
\end{align}
We hence need only to find $u_k$ which satisfies the differential equation \eqref{eq:eom1}.

\subsection{Finding solutions for dislocation displacement gradients}
\label{sec:stroh}
%%%%%%%%%%%%%%%%%%%%%%%%%%%%%%%%%%%%%%%%%

Throughout this work, we limit our analyses to infinite, straight dislocations.
This simplification removes any dependence of the field quantities on the position along the dislocation line.
A. N. Stroh~\cite{Stroh:1962} describes a method to compute solutions for $\vec{u}$ based upon the ansatz
\footnote{
We follow ref.~\cite[pp.~467--478]{Hirth:1982} below.}
\begin{align}
 u_k&=\frac{D A_k}{2\pi i}\ln\left(m_j x_j+p n_j x_j\right)
 \,, \label{eq:disp}
\end{align}
where the perpendicular unit vectors $\vec{m}$ and $\vec{n}$ are conveniently chosen normal to the sense vector $\vec{t}$ of the dislocation whose displacement field we are seeking: i.e. $\vec{t}=\vec{m}\times\vec{n}$.
Inserting this ansatz into the e.o.m. \eqref{eq:eom1} transforms the differential equation into an eigenvalue problem in terms of the unknown coefficients $A_k$ and $p$.
The overall factor $D$ is finally determined by the boundary conditions that a Burgers circuit around the dislocation line has a discontinuity described by the Burgers vector $\vec{b}$, and that there be no external forces present at the dislocation core.

Due to Voigt symmetry, the eigenvalue problem can be formulated in terms of a 6-dimensional vector $\vec\zeta$ and associated $6\times6$ matrix $\mat N$ comprised of four $3\times3$ blocks,
i.e. $\mat N\cdot\vec\zeta=p\vec\zeta$,
see~\cite[pp.~467--473]{Hirth:1982} for details on this ``sextic formalism''.
On the other hand, one can reformulate the theory so that this sextic eigenvalue problem is replaced  by one of evaluating a set of definite integrals~\cite{Barnett:1973,Asaro:1973}.
We will now summarize this (simpler) latter approach in the notation of ref.~\cite{Hirth:1982}.
The crucial observation is that the unit vectors $\vec{m}$, $\vec{n}$ are defined only up to an arbitrary angle $\ph$.
Hence, as one can show, averaging over this angle yields a solution for $u_{j,k}$ in terms of the averaged matrix
\begin{align}
 \langle \mat N\rangle&=\frac1{2\pi}\int\limits_0^{2\pi}\mat N d\phi
 =\begin{pmatrix}
 \mat S & \mat Q \\
 \mat B & \mat S^T
 \end{pmatrix}
\,,
\end{align}
with
\begin{align}
 \mat S&=-\frac1{2\pi}\int\limits_0^{2\pi}(nn)^{-1}(nm)\,\rd\phi  \,, &
\mat S^T&=-\frac1{2\pi}\int\limits_0^{2\pi}(mn)(nn)^{-1}\rd\phi  \,,\nn\\
\mat Q&=-\frac1{2\pi}\int\limits_0^{2\pi}(nn)^{-1}\rd\phi  \,, &
\mat B&=-\frac1{2\pi}\int\limits_0^{2\pi}\left[(mn)(nn)^{-1}(nm)-(mm)\right]\rd\phi
\,, \label{eq:thematrix}
\end{align}
where we have employed the shorthand notation $(ab)_{jk}\coleq a_i \hat C_{ijkl} b_l$.
In particular, the displacement field reads~\cite{Bacon:1980}
\begin{align}
 u_j(\abs{\vec{x}},\phi)&=-\frac{b_l}{2\pi}\left\{S_{jl}\ln\abs{\vec{x}}-S_{il}\int\limits_0^\phi\left[(nn)^{-1}(nm)\right]_{ji}\rd\phi'-B_{il}\int\limits_0^\phi(nn)^{-1}_{ji}\rd\phi'\right\}
 \,. \label{eq:uj-sol}
\end{align}
We emphasize that the displacement field is comprised of the superposition of a radially dependent term ($\ln\abs{\vec{x}}$) and an angular term $\ph$.
Upon selecting a coordinate system which rotates with $\ph$ about the dislocation line ($m_i x_i=r$, $n_ix_i=0$), the displacement gradient computes to~\cite[p.~476]{Hirth:1982}:
\begin{align}
 u_{j,k}(r,\phi)&=\frac{\tilde{u}_{j,k}(\phi)}{r}\,, \nn\\
 \tilde{u}_{j,k}(\phi)&=-\frac{b_l}{2\pi}\left\{m_k S_{jl}-n_k\left[(nn)^{-1}(nm)\right]_{ji}S_{il}-n_k(nn)^{-1}_{ji}B_{il}\right\}
 \,.\label{eq:ukl-sol}
\end{align}
Notice that the $\tilde{u}_{j,k}(\ph)$ dependence on $\ph$ resides within the unit vectors $\vec{m}$, $\vec{n}$.
The tensors $\mat S$, and $\mat B$ are constant projections of the dynamic stiffness tensor onto the plane with unit normal vector $\vec{t}$ as computed from \eqnref{eq:thematrix};
Their values depend on the elastic constants as well as (constant) dislocation velocity, and material density, cf. \eqref{eq:eom1}.
These integrals are directly amenable to numerical integration for any given set of material constants
\footnote{
Analytic results for $\mat S$, $\mat B$ are much more difficult to obtain and are known only in the simplest of cases, namely within the limit of isotropy~\cite{Bacon:1980}, and for dislocations oriented along $\left<110\right>$ within an fcc lattice~\cite{Hirth:1982}.
},
and this numerical integration need be performed only once for a given dislocation velocity, material, and orientation.
The dislocation displacement gradients may subsequently be algebraically assembled according to \eqnref{eq:ukl-sol}.

Note, that this method is limited to the case of steady-state motion leading to the simpler e.o.m. \eqref{eq:eom1}.
Its strength is its simplicity; if one were to study acceleration as well (which is beyond the scope of our present work), a more general method using dynamic Green functions would be necessary~\cite{Markenscoff:2009}.
In the steady state limit, of course, the same dislocation solution is reproduced as with the current method, see e.g.~\cite[p. 53--57]{Mura:1987} for the isotropic case (which we discuss in section~\ref{sec:isotropic} below).

\subsection{Energy and line tension}
\label{sec:linetension}
%%%%%%%%%%%%%%%%%%%%%%%%%%%%%%%%%%%%%%%

The concept of line tension, as applied to a dislocation, follows that of the accumulation of strain energy incurred through undulation of a material surface;
The surface tension is defined as the \emph{change} in energy per unit length due to an infinitesimal undulation of a surface, with respect to a nominally flat surface.
The sign of this \emph{difference} in energy indicates if a wavy configuration of the material surface is energetically favorable.
When considering a dislocation within this context, the line tension is related to both the energy, and the orientation of the dislocation line through the relation \eqnref{eq:def-linetension}, cf.~\cite[pp.~174--177]{Hirth:1982}
% \begin{align}
%  \LT&=E+\frac{\pa^2 E}{\pa \vth^2}
%  \,, \label{eq:def-linetension}
% \end{align}
% which is general, omitting
As noted in the introduction, this relation omits
a constant which depends upon the three dimensional geometry of the instability under examination~\cite{Pueschl:1987}.
% Furthermore, $\vth$ is the angle between line sense and Burgers vector, see figure~\ref{fig:geometry}.

The infinitesimal linear elastic strain energy density~\cite[pp.~33--34]{Hirth:1982}
surrounding the dislocation core follows as $\rd W^s=\s_{ij}\rd\e_{ij}$, or upon integration
\begin{align}
W^s&=\int\s_{ij}\rd\e_{ij}=\frac12C_{ijkl}\e_{ij}\e_{kl}
\,.
\end{align}
Due to Voigt symmetry of the elastic constants one may substitute the infinitesimal strains $\e_{ij}\coleq(u_{i,j}+u_{j,i})/2$ with the displacement gradients.
As before, we emphasize that $u_{i,j}(\vth,\ph)$ depends on the \emph{dynamic stiffness tensor}, as defined earlier, while the elastic strain energy density, as written, depends linearly on the nominal (static) elastic stiffness tensor.
Upon considering finite velocity, in addition to the linear elastic strain energy density $W^s$ we also account for the kinetic energy density due to the temporal evolution of the displacement field,
\begin{align}
 W^k&=\frac{\rho}2\left(\diff{u_i}{t}\right)^2
 =\frac{\rho}2 v_i v_j u_{k,i} u_{k,j}
 \,,
\end{align}
where we have utilized the relation $\dot u_k=-v_j u_{i,j}$.
Hence, the total energy density is
\begin{align}
 W(\vth,v)&=W^s+W^k=\left(C_{ijkl}+\rho v_i \d_{jk} v_l\right) u_{j,i} u_{k,l}/2
 \,,\label{eq:eng_dens}
\end{align}
which depends on both the character of the dislocation and the velocity through the displacement gradients.
Upon integrating the energy density per unit length through a hollow cylinder defined by inner and outer radii $r_0$ and $R$, respectively, the energy per unit length $E$ of a moving dislocation is
\footnote{
% We neglect nonlinearities associated with the dislocation core, since they are expected to remain relatively constant;
By considering \emph{differences} in energy between two similar dislocation configurations, within a sufficiently large crystal the \emph{difference} in linear elastic strain and kinetic energy will always dominate because it scales logarithmically with the crystal radius, $R$.
We hence neglect effects associated with the dislocation core, not only for simplicity but also assuming that the cores remain similar across two configurations and their effects on the line tension is therefore subleading.
For additional details on the energetics of the dislocation core, we refer to, e.g.,~\cite{Clouet:2011a,Clouet:2011b,Luthi:2017,Rodney:2017}.
}
\begin{align}
 E&=\int_{r_0}^{R}r \,\rd r\int_{0}^{2\pi}\rd\ph\, W
  \nn\\&
  =\frac12\left(C_{ijkl}+\rho \d_{jk} v_i(\vth) v_l(\vth)\right) \left[\int_{0}^{2\pi}\rd\ph\, \tilde{u}_{j,i}(\vth) \tilde{u}_{k,l}(\vth)\right]\ln\left(\frac{R}{r_0}\right)
 \,, \label{eq:energy1}
\end{align}
where $\tilde{u}_{i,j}\coleq r u_{i,j}$ is independent of $r$ (see eq. \eqref{eq:ukl-sol}).
Since the only radial dependence resides within $u_{i,j}\propto 1/r$, the integral $u_{i,j}u_{k,l}r\rd r\propto\ln(R/r_0)$ which depends upon an arbitrary constant.
Upon insertion of \eqnref{eq:energy1} into \eqnref{eq:def-linetension} our final expression for the line tension of a steady state moving dislocation within an anisotropic infinite medium is
\begin{align}
 \LT&=\frac12\left(1+\frac{\pa^2 }{\pa \vth^2}\right) \left(C_{ijkl}+\rho \d_{jk} v_i(\vth) v_l(\vth)\right) \left[\int_{0}^{2\pi}\rd\ph\,\tilde{u}_{j,i}(\vth) \tilde{u}_{k,l}(\vth)\right]\ln\left(\frac{R}{r_0}\right)
 \,. \label{eq:linetension}
\end{align}
%%%%
% This is the main result of our work.
We note that our expression for $\Gamma$ does not include an arbitrary constant which is $\mathcal{O}(1)$ in the zero velocity limit.
This is due to our ansatz in eq. \eqref{eq:disp} where we have assumed an infinite, straight dislocation.
We will apply eq. \eqref{eq:linetension} to dislocations of arbitrary character angle contained within the slip plane with subsonic velocities within both isotropic solids as well as in anisotropic fcc and bcc crystals.

\subsection{Crystal lattice dependence and numerical implementation}
\label{sec:implementation}
%%%%%%%%%%%%%%%%%%%%%%%%%%%%%%

\begin{figure}[h!t!b]
 \centering
 \def\svgwidth{0.65\textwidth}
 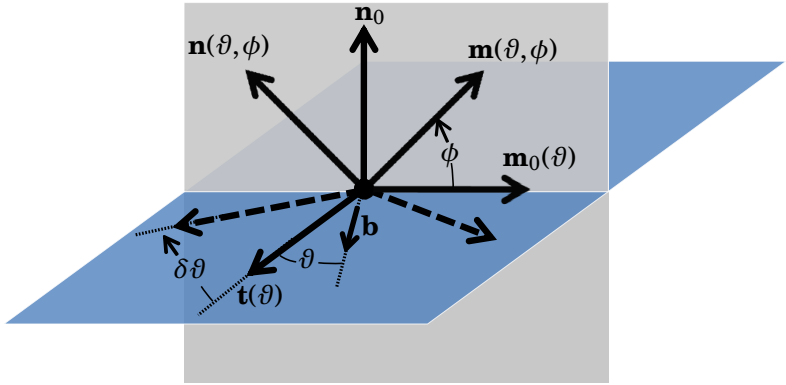
 \caption{Schematic detailing the line direction of dislocation ($\vec{t}$), slip plane normal vector ($\vec{n}_0$), and the two vectors contained within the plane of integration ($\vec{m}$, $\vec{n}$).
 The dislocation line and Burgers vector $\vec{t}$ and $\vec{b}$ lie in the plane shown in blue, whereas the plane spanned by $\vec{m}$ and $\vec{n}$ is shown in gray.}
 \label{fig:geometry}
\end{figure}

Within the previous subsections, we have detailed the computation of the displacement gradient, the dislocation line energy, and dislocation line tension.
Within this section, we relate the two perpendicular vectors ($\mathbf{m}$ and $\mathbf{n}$), which the displacement gradient depends on, to the geometry of the crystalline lattice.
We employ Cartesian coordinates aligned with the crystal axes, and consider a fixed Burgers vector $\vec{b}$ and a unit vector normal to the slip plane under consideration, $\vec{n}_0$.
The material direction of both the Burgers vector and slip plane depend upon the crystal structure.
The sense vector of the dislocation $\vec{t}$ and its dependence on $\vth$ depends upon the Burgers vector and slip plane normal as
\begin{align}
 \vec{t}(\vth)&=\frac1b\left[\vec{b}\cos\vth+\vec{b}\times\vec{n}_0\sin\vth\right]
 \,, \label{eq:def-sensevec}
\end{align}
where $\vth$ is the angle between $\vec{t}$ and $\vec{b}$ and $b\coleq\abs{\vec{b}}$.
Therefore, $\vth=0$ and $\vth=\pi/2$ correspond to pure screw and pure edge dislocations, respectively.
Since we assume that dislocation bow-out will occur within the slip plane spanned by $\vec{t}$ and $\vec{b}$, we may
define a third mutually perpendicular vector $\vec{m}_0(\vth)$ in the direction of motion as
\begin{align}
 \vec{m}_0(\vth)=\vec{n}_0\times \vec{t}(\vth)
 \,.
\end{align}
We will assume, without loss of generality, that the dislocation moves in that direction, i.e. $\vec{v}(\vth)=v\vec{m}_0(\vth)$.
From $\vec{m}_0$, $\vec{n}_0$ we construct
\begin{align}
 \vec{m}(\vth,\ph)&=\vec{m}_0(\vth)\cos\ph + \vec{n}_0\sin\ph\,, \nn\\
 \vec{n}(\vth,\ph)&=\vec{n}_0\cos\ph - \vec{m}_0(\vth)\sin\ph\,,
\end{align}
which span the plane normal to $\vec{t}$, as illustrated within figure~\ref{fig:geometry}.

In order to evaluate eq. \eqref{eq:linetension} we must resort to numerical methods, since it is analytically insoluble.
We have implemented eq. \eqref{eq:linetension} within a numerical code
\footnote{
A second suite of computations of \eqref{eq:linetension} was performed within Mathematica
for validation purposes.
Within this implementation, differentiation w.r.t. $\vth$ was performed symbolically and only the integrations over $\ph$ were completed numerically.
}
which relies upon the stiffness tensor ($C_{ijkl}$), the Burgers vector direction ($\vec{b}/|\vec{b}|$), and the slip plane normal ($\vec{n}_0$) as input.
The integrals over $\ph$ (i.e. $\tilde{u}_{i,j}$, $B_{ij}$, etc.) are performed using a trapezoidal numerical integrator with a step size of $\delta \ph \approx \pi/500$;
The derivatives with respect to $\vth$ are evaluated by successively differencing evaluations of $E(\vth, v)$
(\eqnref{eq:energy1})
over small increments of $\delta \vth$ (e.g. $\frac{\pa E(\vth,v)}{\pa \vth} \coleq (E(\vth+\delta \vth, v) - E(\vth-\delta \vth, v)) /2\delta \vth$), where $\delta \vth \approx \pi/600$.
For each of the materials investigated within this study, the resulting $\LT$ has been evaluated at roughly 600 evenly spaced intervals of $\vth$ where $-\pi/2 < \vth <
\pi/2$, and 625 evenly spaced intervals of $\bm\coleq\sqrt{{\rho v^2}/{\bar{\mu}}}$ within the range $0 \leq \bm \leq 1$, with $\bar\mu\coleq\frac14\left(c_{11}-c_{12}+2c_{44}\right)$ denoting the mean shear modulus.
This results in $\sim 375,000$ data points for each material.
Throughout the remainder of this work, it will be convenient to show and discuss results in the rescaled, dimensionless line tension prelogarithmic factor $\lt$ via
\begin{align}
 \LT &=\left(\frac{\bar\mu b^2}{4\pi}\ln\frac{R}{r_0}\right)\lt
 \,. \label{eq:linetensionprefactor}
\end{align}
Our results will be presented in terms of $\lt(\vth,\bm(v))$ (for which we will use the terms ``line tension'' and ``prelogarithmic factor'' interchangeably) as a function of dislocation character $\vth$ and dimensionless, scaled velocity $\bm$.

\section{The isotropic limit}
\label{sec:isotropic}
%%%%%%%%%%%%%%%%%%%%%%%%%%%%%%%

We begin by examining the energy, and line tension of a moving dislocation in the isotropic limit.
In this limit, the expressions for the energy and line tension are considerably simplified and are analytically tractable, yielding insight towards our anisotropic results.
Referring to eq. \eqref{eq:STIFF}, in the limit that $c_{11}\to c_{12}+2c_{44}$ there is no orientational dependence dictated by the crystal symmetry;
the dislocation displacement gradients, energy, and line tension do not depend upon the orientation of the dislocation within the crystal.
Upon adopting a coordinate system fixed to the dislocation line with $\vec{t}$ oriented along $\hat z$ and $\vec{n}_0$ oriented along $\hat y$, the $\vth$ dependence resides entirely within the Burgers vector, and the vectors defined within section~\ref{sec:method} reduce to
\begin{align}
 \vec{t}&=(0,0,1)\,,& 
\vec{m}&=\left(\cos\ph,\sin\ph,0\right)\,,& \vec{n}&=\left(-\sin\ph,\cos\ph,0\right)\,,\nn\\
 \vec{v}&=(v,0,0) \,, &
 \vec{b}&=(b\sin\vth,0,b\cos\vth)\,, &
 r&=\vec{m}\cdot\vec{x}=\sqrt{x^2+y^2}
 \,.
\end{align}
One may verify that, in these coordinates, the well known solutions for stationary moving edge and screw dislocations within an isotropic crystal~\cite{Eshelby:1949,Weertman:1961,Weertman:1980} are reproduced: 
\begin{align}
 u_{x,x}&=\frac{-b_e y}{\pi  \bt ^2 }\left(\frac{1/\gl }{\left((x-t v)^2 +y^2/\gl ^2\right)}-\frac{\left(1-\frac{\bt ^2}{2}\right) /\gt }{\left((x-t v)^2 +y^2/\gt ^2\right)}\right)
 \,,\nn\\
 u_{x,y}&=\frac{b_e (x-t v)}{\pi  \bt ^2 }\left(\frac{1/ \gl }{ \left((x-t v)^2 +y^2/\gl ^2\right)}-\frac{\left(1-\frac{\bt ^2}{2}\right)/ \gt }{\left((x-t v)^2 +y^2/\gt ^2\right)}\right)
 \,,\nn\\
 u_{y,x}&=\frac{b_e (x-t v)}{\pi  \bt ^2 }\left( \frac{1/\gl }{\left((x-t v)^2 +y^2/\gl ^2\right)}-\frac{\gt \left(1-\frac{\bt ^2}{2}\right) }{ \left((x-t v)^2 +y^2/\gt ^2\right)}\right)
 \,,\nn\\
 u_{y,y}&=\frac{b_e y}{\pi  \bt ^2 }\left(\frac{1/\gl ^3}{ \left((x-t v)^2 +y^2/\gl ^2\right)}-\frac{\left(1-\frac{\bt ^2}{2}\right)/ \gt }{\left((x-t v)^2 +y^2/\gt ^2\right)}\right)
 \,,\nn\\
 u_{z,x}&=-\frac{b_s\, y/\gt }{2 \pi  \left((x-t v)^2+y^2/\gt^2\right)}\,, \qquad\qquad
 u_{z,y}=\frac{b_s\, (x-t v)/ \gt}{2 \pi  \left((x-t v)^2+y^2/\gt^2 \right)}
 \,, \label{eq:u-isotropic}
\end{align}
where we have employed a dislocation fixed coordinate system ($x\to(x-vt)$) and where
\footnote{Note that our definition of $\g$ matches the one known from special relativity and is related to Weertman's~\cite{Weertman:1980} notation via $\b_\txt{W}\rightsquigarrow 1/\g$, $a_\txt{W}\rightsquigarrow c$.}
\begin{align}
 \gt&=\frac{1}{\sqrt{1-\bt^2}}\,, &
 \bt&=\frac{v}{\ct}\,,
 &
 \gl&=\frac{1}{\sqrt{1-\bl^2}}\,, &
 \bl&=\frac{v}{\cl}
 \,,
\end{align}
and $\ct=\sqrt{\m/\rho}$, $\cl=\sqrt{\left(\l+2\m\right)/\rho}$ are the transverse and longitudinal sound speed, respectively.
The edge and screw components of the Burgers vector, $b_e\coleq b\sin\vth$ and $b_s\coleq b\cos\vth$, are the only variables depending on the angle $\vth$.

\subsection{Energy of a moving dislocation in an isotropic crystal}
%%%%%%%%%%%%%%%%%%%%%%%%%%%%%%%%%%%%%%%%%%%

It is readily verified that even for finite velocities, the edge and screw components of dislocations within isotropic media do not interact energetically, and therefore both the displacements, and resulting energies of an arbitrarily oriented dislocation may be superposed.
Specifically, since $C_{ijkl}u_{i,j}^\txt{edge}u_{k,l}^\txt{screw} = 0$, there is no cross term in the total energy of a mixed dislocation and we are justified superposing the energies of edge and screw components.

We note also that $\e_{xy}^\txt{edge}=(u^\txt{edge}_{x,y}+u^\txt{edge}_{y,x})/2$ vanishes at $y=0$ (and hence changes sign across the glide plane) at a dislocation velocity $v=\cRl$ which is known as the Rayleigh wave velocity (i.e. the velocity of surface waves)~\cite{Weertman:1961,Weertman:1980}.
It is defined by the relation $\gl \gt =(1-\bt ^2/2)^{-2}$ and is always smaller than (but fairly close to) the transverse sound speed, $\cRl<\ct $,
a typical value being $\cRl\approx 0.93\ct $ if Poisson's ratio is $1/3$ (corresponding to $\cl=2\ct$).

An additional simplification occurs in the strain energy of the screw dislocation:
$W^s=\frac{\l}{2}(\e_{ii})^2+\m\e_{ij}\e_{ij}$,
and since the strain of the screw dislocation is traceless, the dependence on $\l$ will only appear for the edge dislocation, i.e.
\begin{align}
 \e_{ii}^\txt{screw} & = 0\,, &
 \e_{ii}^\txt{edge} &=\frac{-b\,\ct^2 y}{\pi  \cl ^2 \gl\left((x-t v)^2 +y^2/\gl ^2\right)}
 \,,
\end{align}
see eqs. \eqref{eq:Hooke}, \eqref{eq:u-isotropic}.
The total energy for an arbitrary dislocation orientation follows as
\begin{align}
 E^\txt{mixed}(\vth,v)&= E_0\left\{\gt \cos^2\vth +\frac{\sin^2\vth}{\bt^2}\left[\frac{8}{\gl}+4\gl+\gt^3\left(1-\frac{6}{\gt^2}-\frac{7}{\gt^4}\right)\right]\right\}
 \,, \nn\\
 E_0&=\frac{\m b^2}{4\pi}\ln\left(\frac{R}{r_0}\right)
 \,, \label{eq:energy-mixed}
\end{align}
with the contributions from screw and edge dislocations scaled by $\cos^2\vth$ and $\sin^2\vth$, respectively (see also~\cite{Weertman:1980}).

In the limit $v\to0$ the energies for both edge and screw dislocations reduce to the well-known expressions~\cite{Weertman:1961,Lothe:1967}
\begin{align}
 E^\txt{screw}(0,0)&=E_0 \,, &
 E^\txt{edge}(\pi/2,0)&=E_0\frac{2(\l+\m)}{\l+2\m}
%  = E_0\frac{2(\cl^2-\ct^2)}{\cl^2}
=\frac{E_0}{1-\n}
\,,
\end{align}
where Poisson's ratio is related to the Lam{\'e} constants via $\n=\l/2(\m+\l)%=\frac{\frac12\cl^2-\ct^2}{\cl^2-\ct^2}
$.
On the other hand, as the velocity approaches the transverse sound speed, both energies diverge as
\begin{align}
 \lim\limits_{v\to\ct}E^\txt{screw}(0,v)&\to\frac{E_0}{\sqrt{1-\bt^2}}
 \,,\nn\\
 \lim\limits_{v\to\ct}E^\txt{edge}(\pi/2,v)&\to \frac{E_0}{2\sqrt{2}}\left(\frac{1}{(1-\bt^2)^{\frac32}}+\frac{37}{4\sqrt{1-\bt^2}}+\frac{16 (2-\nu)}{\sqrt{1-\nu }}+\cO\left(\sqrt{1-\bt^2}\right)\right)
 \,. \label{eq:isotropic-energy-divergences}
\end{align}
All divergences appearing in the energy as well as the displacement gradients as $v\to\ct$ can be traced back to the expression $(nn)^{-1}$ appearing in \eqref{eq:thematrix} and \eqref{eq:ukl-sol}, and are due to
\begin{align}
 \det\left(nn\right)&=\det\left(\vec{n}\cdot\hat{\mat{C}}\cdot\vec{n}\right)
 =\det\left(\vec{n}\cdot\mat{C}\cdot\vec{n}-\rho \left(\vec{n}\cdot\vec{v}\right)^2\id\right)
 \nn\\
 &=0
 \,,
\end{align}
as $v\to\ct$ (or $v\to\cl$) and $\ph\to\pm\pi/2$ in which case $\vec{n}\parallel\vec{v}$.
In fact, the leading divergence as $v\to\ct$ and $\ph\to\pm\pi/2$ is $(nn)^{-1}\sim 1/(1-\bt^2)$.
Due to averaging over $\ph$, the matrices $\mat{S}$, $\mat{B}$ exhibit a weaker divergence of $1/\sqrt{1-\bt^2}$.
Upon inspecting expression \eqref{eq:ukl-sol}, it is clear that the edge dislocation has a divergence of order $1/(1-\bt^2)^{3/2}$ in the displacement gradient ($u_{i,j}$) originating from the collection of terms $-n_k\left[(nn)^{-1}(nm)\right]_{ji}S_{il}-n_k(nn)^{-1}_{ji}B_{il}$.
In the solution for pure screw dislocations, two types of simplification lead to the weaker divergence of $1/\sqrt{1-\bt^2}$ in the displacement gradient:
Firstly, $\mat{S}\cdot\vec{b}=0$ for geometrical reasons, and furthermore $\lim\limits_{v\to\ct}\mat{B}\cdot\vec{b}\propto\sqrt{1-\bt^2}\to0$.
Thus, all but the last term in \eqref{eq:ukl-sol} vanish in the pure screw dislocation case, and that surviving term, $n_k(nn)^{-1}_{ji}B_{il}$, furthermore exhibits a weaker divergence than in the edge case.
We emphasize that the leading divergences in $u_{i,j}$ outlined above occur only at particular angles, $\ph=\pm\pi/2$, i.e. perpendicularly to the direction of dislocation motion.
This is also the reason why the degree of divergence in the dislocation energy $E$ (which involves an integration over $\phi$) is not twice that of $u_{i,j}$, despite $u_{i,j}$ entering quadratically.

Within this work, we limit our analysis to the subsonic regime.
Solutions for the strain fields of straight dislocations have also been derived for the transonic and supersonic regimes~\cite{Eshelby:1956,Weertman:1980}.
All of these solutions, however, diverge at $\ct$ and $\cl$, i.e. they cannot describe the vicinity of the two sound speeds.
For more recent literature on supersonic dislocations, see~\cite{Markenscoff:2009,Pellegrini:2010,Barnett:2002,Pellegrini:2017} and references therein.

\subsection{Line tension and stability}
%%%%%%%%%%%%%%%%%%%%%%%%%%%%%%%%%%%

\begin{figure}[h!t!b]
 \centering
 \includegraphics[width=0.6\textwidth]{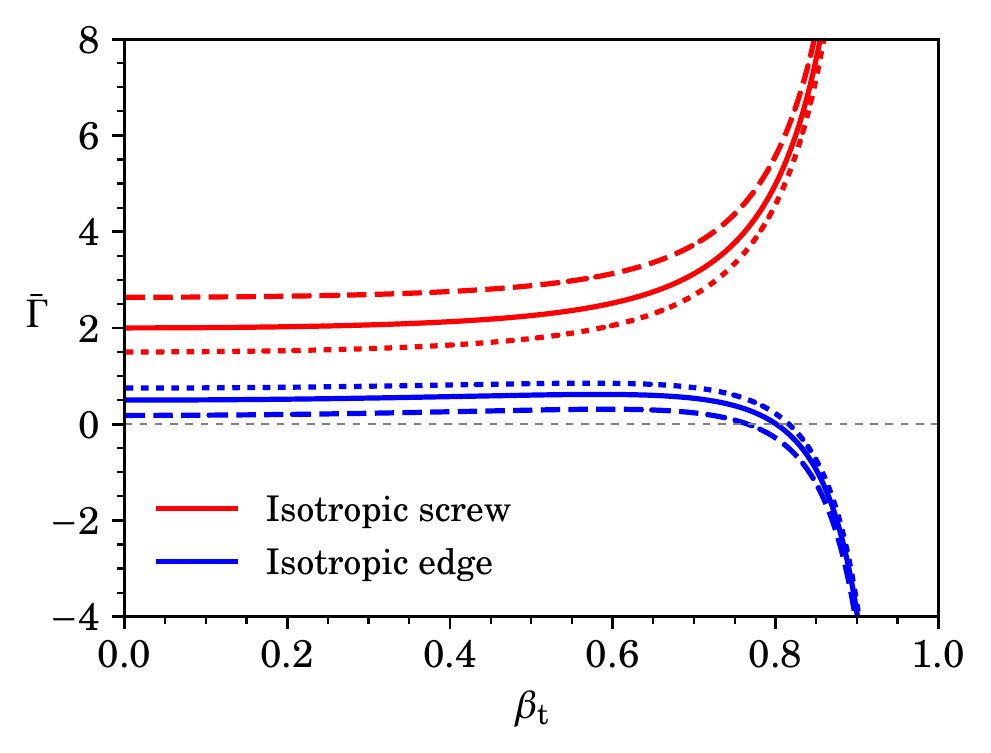}
 \caption{%
 The line tension ($\lt$) versus the dimensionless velocity ($\beta_t=v/c_t$) is depicted for both screw (red) and edge (blue) dislocations in the isotropic limit (i.e. both figure axes are in dimensionless units).
 The solid lines (--) depict the line tension for Poisson's ratio equal to $\nu=1/3$.
 For the dashed lines (- -) we have selected $\nu=0.45$ and for the dotted lines ($\cdot\cdot$) $\nu=0.2$.}
 \label{fig:linetension-isotropic}
\end{figure}

The line tension prelogarithmic factor ($\lt$) follows from application of \eqnref{eq:def-linetension} to $E(\vth,v)$,
using our isotropic solution (\eqnref{eq:energy-mixed}) for $E$ (see ref.~\cite{Hirth:1982} for the static case).
We immediately find
\begin{align}
 \LT(\vth,v)&= \frac{E_0}{\bt^2}\left\{2\cos^2\vth\left[\frac{8}{\gl}+4\gl+\gt^3\left(1-\frac{6}{\gt^2}-\frac{7}{\gt^4}\right)-\frac{\bt^2\gt}{2}\right]\right.\nn\\
 &\quad\qquad \left. -\sin^2\vth\left[\frac{8}{\gl}+4\gl+\gt^3\left(1-\frac{6}{\gt^2}-\frac{7}{\gt^4}\right)-2\bt^2\gt\right]\right\}
 .\label{eq:linetension-isotropic}
\end{align}
In the zero velocity limit this expression tends to the classical result given within ref.~\cite[p.~176]{Hirth:1982}:
\begin{align}
 \lim\limits_{v\to0}\LT(\vth,v)&= \frac{E_0}{(1-\n)}\Big\{(1+\n)\cos^2\vth+(1-2\n)\sin^2\vth\Big\}
 ,
\end{align}
where $\n$ denotes Poisson's ratio.
For metals of interest, $\n\sim1/3$, which indicates that the line tension of a static screw dislocation is roughly four times larger than for a static edge dislocation, i.e.:
$\LT(\vth,v=0,\n=1/3)=E_0\left(2\cos^2\vth+\frac12\sin^2\vth\right)$.
Figure~\ref{fig:linetension-isotropic} shows the velocity dependence of the line tension $\lt(0,\bt \ct)$ in red and $\lt(\pi/2,\bt \ct)$ in blue, where, referring to \eqnref{eq:linetensionprefactor}
\begin{align}
 \lt(\vth,v)&\coleq \frac{\LT(\vth,v)}{E_0}
 \,,
\end{align}
is the dimensionless, rescaled line tension prelogarithmic factor in the isotropic limit.

What is particularly interesting about the full velocity dependent expression \eqref{eq:linetension-isotropic}, is that the divergent term $\gt^3$ is negative in the pure edge case.
This means that the line tension of an edge dislocation (where $\vth=\pi/2$) can become zero (in contrast to the screw case) indicating that it will cease to be stable in a straight configuration.
In particular, for $\n\approx1/3$ we find that the line tension of an edge dislocation becomes zero at around $80\%$ transverse sound speed, whereas it remains positive everywhere in the pure screw case --- see figure~\ref{fig:linetension-isotropic}.
The range of velocities where this happens, depending on the value of Poisson's ratio (which lies in the range $-1<\n<1/2$),
is roughly $0.72\lesssim\bt^0\lesssim0.83$; this is always beneath the Rayleigh speed.
Finally, upon considering specific geometries of bow-out, a $\vth$-dependent ``constant'' ($\Gamma_0$) may need to be superposed to the expression for the line tension \eqref{eq:def-linetension} (i.e. corresponding to different constants for edge and screw dislocations), see e.g. ref.~\cite[pp.~176--177]{Hirth:1982}.
This addition would only change the value of the velocity $\bt^0$, but not the main result:
that $\bt^0<\ct$ exists only for edge dislocations and is smaller than (but fairly close to) the transverse sound speed.

This observation is in agreement with previous MD simulation results~\cite{Marian:2006,Daphalapurkar:2014,Oren:2017} which indicate a sudden jump in velocity versus stress to the transonic regime for edge dislocations, whereas the transition is smooth for screw dislocations.

It is also instructive to analyze at which angle $\vth$ a mixed dislocation in the isotropic limit becomes unstable at high velocities.
The change in asymptotic behavior from positive to negative occurs when the coefficient of $\g_t^3$ becomes zero in \eqnref{eq:linetension-isotropic}, i.e. when $\vth=\arctan\sqrt{2}$ (independent of Poisson's ratio).
Figure~\ref{fig:linetension-fcc_contour} (a) below shows the regions of stability of such a mixed dislocation in the form $\lt(\vth,v)$.
For the purpose of this figure we have chosen $\n=1/3$.
Other values of $\n$ will change the contours within the low to intermediate velocity regime, but not in the asymptotic region $\bt\to1$, since the divergent terms do not depend on $\n$ (see \eqnref{eq:isotropic-energy-divergences}).

Finally, we need to comment on earlier, similar works that have analyzed the stability of dislocations at high velocities:
Our result disagrees with the earlier assessment of ref.~\cite{Beltz:1968} (see also~\cite{Malen:1970}), whose author found that screw dislocations become unstable at high velocities whereas edge dislocations remain stable.
On the other hand, our results are in perfect agreement with ref.~\cite{Sakamoto:1991}.
A somewhat different stability analysis was performed in the anisotropic case for cubic metals in ref.~\cite{Teutonico:1962}.
The authors' stability analysis was based upon finding the velocity where the force of interaction between two parallel dislocations on the same slip plane changes sign, and the result is that this happens at some velocity lower than the shear wave speed for both edge and screw dislocations.

Within the next section, we extend the general stability analysis ($\lt(\vth,v)$) of dislocations to anisotropic crystals; We concentrate on fcc and bcc crystals.

\section{Cubic crystals}
\label{sec:cubic}
%%%%%%%%%%%%%%%%%%%%%%%%%%%%%

The isotropic analyses and results presented within the previous section provide insight due to their simple, analytical form (i.e., \eqnref{eq:linetension-isotropic}).
Since realistic material crystals are anisotropic however, the approximated isotropic computations have the capacity to obscure subtle details from our line tension calculations.
Within this section, we compute and present results for $\lt$ (\eqnref{eq:linetensionprefactor}) employing full elastic anisotropy with the elastic constants found within table~\ref{tab:values-metals}.

\subsection{Slip systems of dislocations and divergences}

The atomic structure of fcc and bcc crystals dictates the slip plane normal and Burgers vector with respect to reference cubic crystal axes,
\begin{align}
 \vec{b}^\txt{fcc}&=\frac{b}{\sqrt{2}}\left(1,1,0\right)\,, &
 \vec{n}_0^\txt{fcc}&=\frac{1}{\sqrt{3}}\left(1,-1,1\right)
 \,, \nn\\
 \vec{b}^\txt{bcc}&=\frac{b}{\sqrt{3}}\left(1,-1,1\right)\,, &
 \vec{n}_0^\txt{bcc}&=\frac{1}{\sqrt{2}}\left(1,1,0\right)
 \,. \label{eq:slipplanes-bcc}
\end{align}
Following section~\ref{sec:implementation} we may construct both the dislocation line direction ($\vec{t}$) and velocity vector (v$\vec{m}_0$) within the plane of slip.
The velocity vector with respect to the moving dislocation depends on $\vth$ as
\begin{align}
 \vec{v}^\txt{fcc}(\vth)&=v\vec{m}_0^\txt{fcc}
%  =\frac{v}{b}\vec{n}_0^\txt{fcc}\times \left[\vec{b}^\txt{fcc}\cos\vth+\vec{b}^\txt{fcc}\times\vec{n}_0^\txt{fcc}\sin\vth\right]
 =\frac{v}{\sqrt{6}}\left(\sqrt{3} \sin (\vth )-\cos (\vth ),\sqrt{3} \sin (\vth )+\cos (\vth ),2 \cos (\vth )\right)
 \,, \nn\\
 \vec{v}^\txt{bcc}(\vth)&=v\vec{m}_0^\txt{bcc}
 =\frac{v}{\sqrt{6}}\left(\sqrt{2} \sin (\vth )+\cos (\vth ),-\sqrt{2} \sin (\vth )-\cos (\vth ),\sqrt{2} \sin (\vth )-2 \cos (\vth )\right)
 \,.
\end{align}
Considering the two limits, we have
$\vec{v}^\txt{fcc}(0)=\frac{v}{\sqrt{6}}\left(-1,1,2\right)$, $\vec{v}^\txt{bcc}(0)=\frac{v}{\sqrt{6}}\left(1,-1,-2\right)$
for screw dislocations, and
$\vec{v}^\txt{fcc}(\pi/2)=\frac{v}{\sqrt{2}}\left(1,1,0\right)$, $\vec{v}^\txt{bcc}(\pi/2)=\frac{v}{\sqrt{3}}\left(1,-1,1\right)$
for edge dislocations.

Before proceeding immediately to the dependence of the line tension on the velocity, crystal structure, and character angle, an important subtlety which requires consideration within anisotropic media is divergences in dislocation energy at high velocities.
As in the isotropic limit, these divergences occur at the lowest critical velocity leading to $\det(nn)=0$ at certain angle(s) $\ph$.
However, this critical velocity $v_\txt{crit}$ is \emph{not} necessarily the lowest transverse sound speed of the crystal, nor is it the lowest transverse sound speed with respect to the direction of dislocation movement, however in between the two.
To demonstrate this point, the lowest transverse sound speed in the direction of $\vec{v}$ is computed from
\begin{align}
 \det\left(\vec{v}\cdot\mat{C}\cdot\vec{v}-\rho v^2\id\right)\Big|_{v=v_{\txt{t}+}}=0
 \,, \label{eq:sound1}
\end{align}
and we denote this velocity $v_{\txt{t}+}$ because due to $(\vec{v}\cdot\vec{n})^2\le v^2$ (as we will see shortly) it constitutes an upper bound on the critical velocity responsible for divergences in both the dislocation energy and line tension.
Within the plane spanned by $\vec{n}$ as $\ph$ spans the range $0\to2\pi$, there is an angle $\ph_-$ which yields the lowest transverse sound speed $v_{\txt{t}-}$ within the plane of the dislocation.
At $\ph_-$ the relation
\begin{align}
 \det\left(\vec{n}\cdot\mat{C}\cdot\vec{n}-\rho v^2\id\right)\Big|_{\ph=\ph_-,\,v=v_{\txt{t}-}}=0
 \label{eq:sound2}
\end{align}
is satisfied.
Eqs. \eqref{eq:sound1} and \eqref{eq:sound2} provide an upper and lower bound, respectively, for the critical velocity of the divergence.
At a crystal dependent critical angle, $\ph_{\txt{crit}}$, within the plane of the dislocation at critical velocity $v_{\txt{crit}}$ the following relation is satisfied
\begin{align}
 \det\left(\vec{n}\cdot\mat{C}\cdot\vec{n}-\rho \left(\vec{n}\cdot\vec{v}\right)^2\id\right)\Big|_{\ph=\ph_\txt{crit},\,v=v_\txt{crit}}=0
 \,. \label{eq:def-vcrit}
\end{align}
Since $(\vec{v}\cdot\vec{n})^2\le v^2$ (where the equal sign corresponds to $\ph=\pm\pi/2$), we always have
\begin{align}
 v_{\txt{t}-}\le v_\txt{crit} \le v_{\txt{t}+}
 \,,
\end{align}
and in the isotropic limit all three velocities coincide because there is only one shear modulus.
We emphasize that $\ph_{\txt{crit}}$, the angle corresponding to the critical velocity is material dependent.
Only $\ph_-$ depends only on the crystal structure and may be determined to be $\ph_-^\txt{fcc}=\arcsin(1/\sqrt{3})$ and $\ph_-^\txt{bcc}=\pi/3$ for fcc and bcc crystals, respectively.

\begin{table}[h!t!b]
\centering
\caption{\label{tab:velocities}We list the relevant computed critical velocities for the metals investigated within this study.
We have introduced the abbreviated notation $v_{\bar\m}$ to denote a ``velocity'' calculated with respect to the mean shear modulus which is employed within our figures as a scaling parameter.
For the fcc metals, $v_\txt{t}^{\bar{1}12}$, and $v_\txt{t}^{110}$ denote the lowest shear wave speeds for the $(\bar{1}12)$, $(110)$ directions, which correspond to the directions of motion for screw and edge dislocations, respectively.
For the bcc metals, $v_\txt{t}^{1\bar{1}\bar{2}}$, and $v_\txt{t}^{1\bar{1}1}$ denote the lowest shear wave speeds for the $(1\bar{1}\bar{2})$, $(1\bar{1}1)$ directions, which correspond to the directions of motion for screw and edge dislocations, respectively.
In addition, we list the critical velocities, $v_{\txt{crit}}$, at which the line tension diverges for each crystal structure and dislocation orientation within this study.}

\begin{tabular}{lccccc}
\hline
\hline
\vspace{-0.42cm}\\
\vspace{0.1cm}
fcc	&	$v_{\bar\m}$[m/s]		&	$v_\txt{t}^{\bar{1}12}$[m/s]		&	$v_\txt{crit}^{\bar{1}12}$[m/s]	&	$v_\txt{t}^{110}$[m/s]	&	$v_\txt{crit}^{110}$[m/s]	\\
\hline
Al	&	3089	&	3032	&	3024	&	2929	&	2929 \\
Cu	&	2348	&	2040	&	1994	&	1606	&	1606 \\
Ni	&	3098	&	2750	&	2704	&	2288	&	2288 \\
\hline
\vspace{-0.42cm}\\
\vspace{0.1cm}
bcc	&	$v_{\bar\m}$[m/s]		&	$v_\txt{t}^{1\bar{1}\bar{2}}$[m/s]		&	$v_\txt{crit}^{1\bar{1}\bar{2}}$[m/s]	&	$v_\txt{t}^{1\bar{1}1}$[m/s]	&	$v_\txt{crit}^{1\bar{1}1}$[m/s]	\\
\hline
Fe	&	3178	&	2814	&	2585	&	2925	&	2745 \\
Nb	&	2223	&	2101	&	1997	&	2340	&	2147 \\
Ta	&	2032	&	1943	&	1918	&	1957	&	1926 \\
W	&	2877	&	2875	&	2875	&	2875	&	2875 \\
\hline
\hline
\end{tabular}
\end{table}

Within table~\ref{tab:velocities} we list explicit values for $v_{\txt{t}+}$ and $v_\txt{crit}$ for the various fcc and bcc metals investigated within this study computed using the densities and elastic constants listed in table~\ref{tab:values-metals}.
Since the velocity vectors are aligned along different directions for screw and edge dislocations, values for both cases are listed separately.
Notice that only for edge dislocations in fcc crystals, the critical velocity happens to equal the lowest shear wave speed corresponding to the direction of dislocation motion, i.e. $v_\txt{crit}^{110}=v_\txt{t}^{110}$ (see table~\ref{tab:velocities}).
Finally, in order to have a single scaling parameter within all of our plots, we introduce a ``velocity'' calculated with respect to the mean shear modulus ($\bar{\mu}\coleq(c_{11}-c_{12}+2c_{44})/4$), $v_{\bar\m}$ which is listed for all cases as well for comparison.

\subsection{Line tension of pure screw and edge dislocations}

\begin{figure}[h!t!b]
 \centering
 \includegraphics[width=\textwidth]{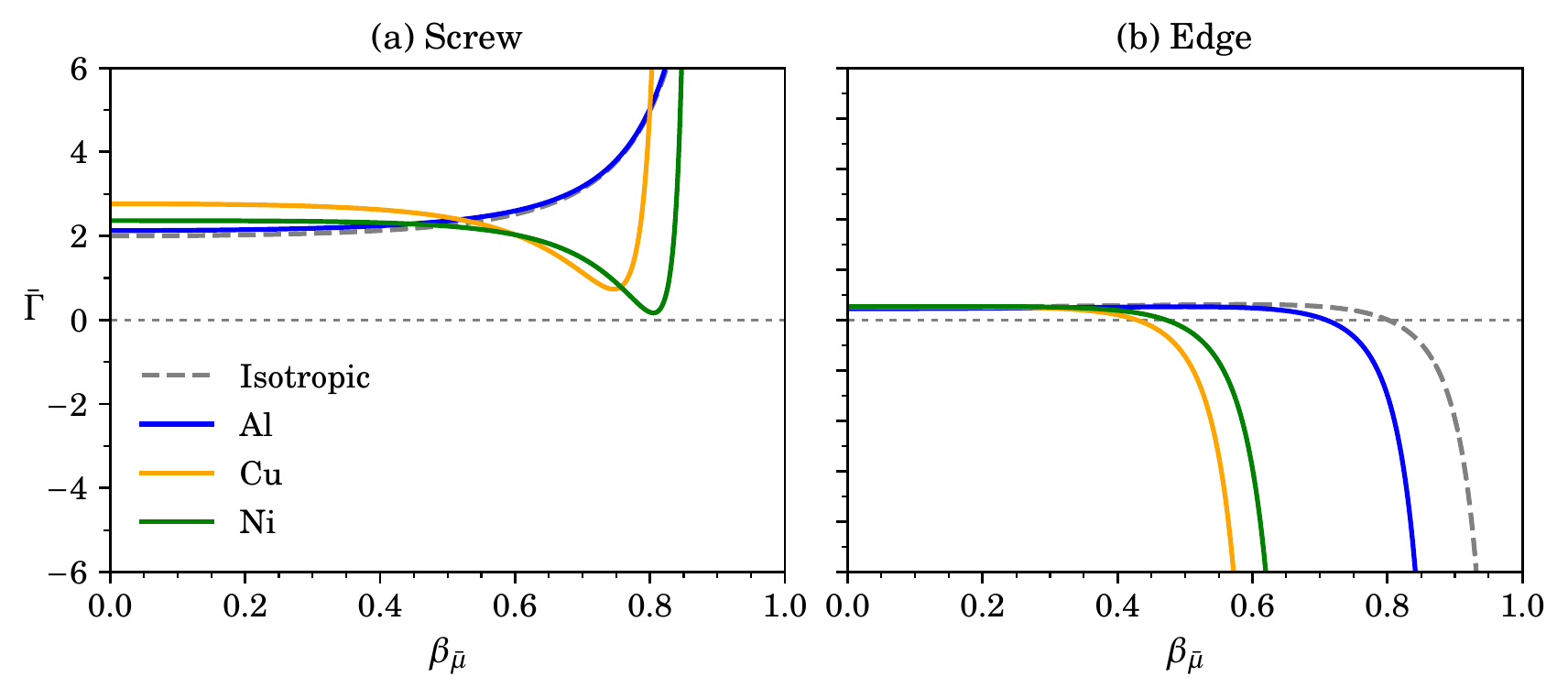}
 \caption{%
 The line tension prelogarithmic factor ($\lt$) for (a) screw and (b) edge dislocations within various fcc metals is shown as a function of $\bm=v\sqrt{\rho/\bar\m}$, and compared to the isotropic limit (dashed gray curve).}
 \label{fig:linetension-fcc}
\end{figure}

\noindent
Using the elastic constants listed within table~\ref{tab:values-metals} as well as the Burgers vector and slip plane normal specific to fcc and bcc crystal structures, we have computed the line tension for a selection of fcc and bcc metals from \eqnref{eq:linetensionprefactor} with \eqref{eq:linetension}, as described in section~\ref{sec:linetension}.
Figure~\ref{fig:linetension-fcc} illustrates the line tension of three fcc metals (Al, Cu, Ni) versus dimensionless velocity ($\bm$), as compared to the isotropic limit with Poisson's ratio $\n=1/3$.
We begin with directing our attention to the overall behavior of the curves.
As compared to elastic isotropy, one sees relatively similar behavior in that screw dislocations remain stable up to high velocities whereas edge dislocations reach instability at some critical velocity (which might be shifted by a finite but unknown constant due to $\Gamma_0$ within \eqref{eq:def-bowout}).
The main differences for increasingly anisotropic fcc metals are the $\bm$ of the divergences, which decrease with increasing anisotropy.
Additionally, screw dislocations exhibit a moderate drop in line tension close to $\bm\approx0.8$ before becoming infinitely stable at their critical velocity, $v_\txt{crit}$.

Focusing our attention specifically on figure~\ref{fig:linetension-fcc} (a), Al (in blue) behaves very similar to the isotropic result which is as to be expected, and similar to figure~\ref{fig:linetension-isotropic}.
With increasing anisotropy, an additional competing divergence (which is negative) becomes evident, which is both weaker and of opposite sign than the positive divergence present at the critical velocity.
Since there is an additional arbitrary constant~\cite{Pueschl:1987} which we have not included within this work, it is possible that both Ni and Cu exhibit negative line tensions over a range of $0.7\leq\bm\leq0.8$.
However, comparisons to MD simulation results~\cite{Daphalapurkar:2014,Marian:2006,Oren:2017}, suggest that this is not the case, and the constant is likely either small or positive.

Within figure \ref{fig:linetension-fcc} (b) both our isotropic analyses, and our three select anisotropic fcc metals exhibit a negative divergence at a fraction of the scaled velocity i.e. $0.7\leq \bm \leq 0.95$, all of which is less than our isotropic example which diverges at $\bm=1$.
Therefore, with the inclusion of and increasing degree of anisotropy the predicted velocity of the instability for nominally fcc edge dislocations decreases.

\begin{figure}[h!t!b]
 \centering
 \includegraphics[width=\textwidth]{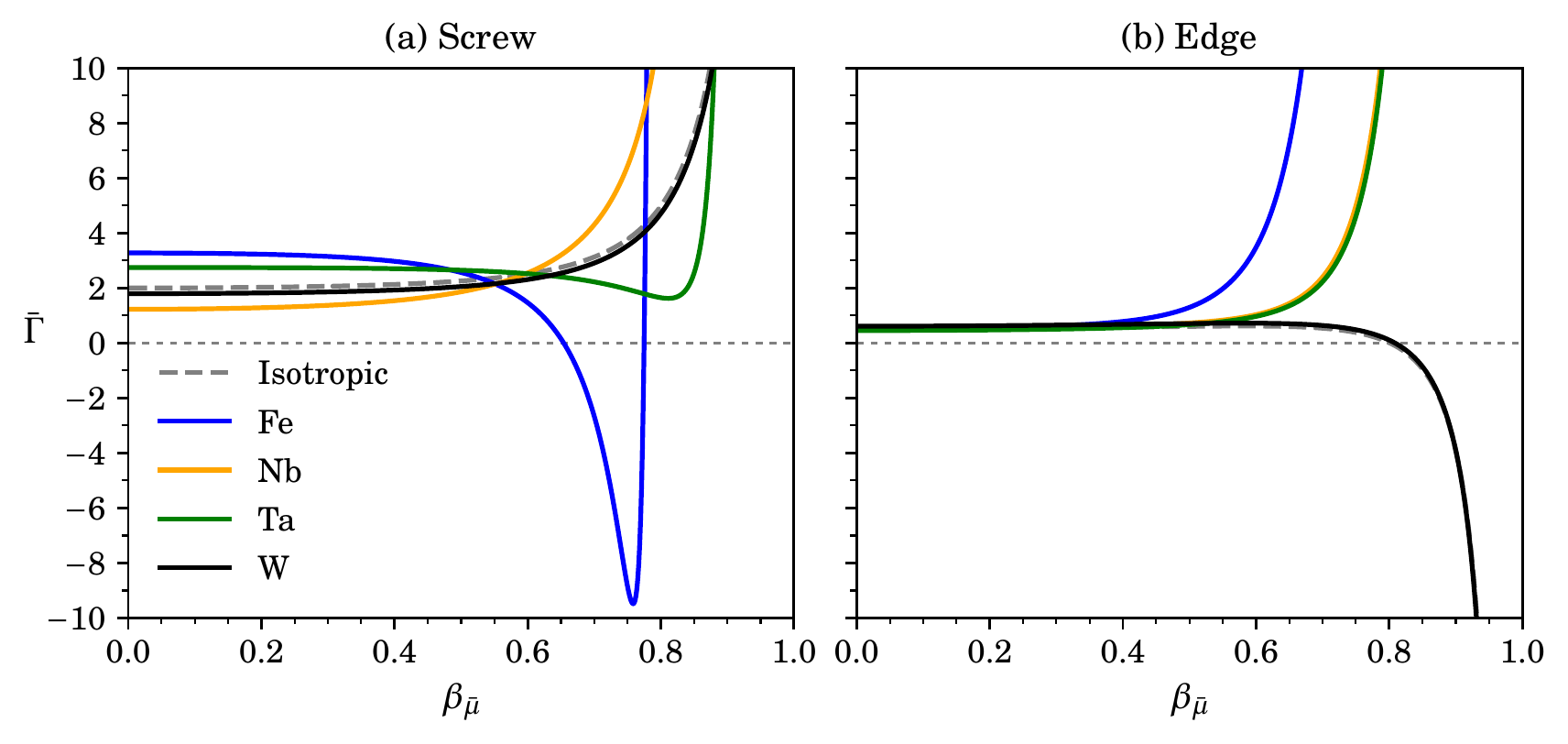}
 \caption{%
 The line tension prelogarithmic factor ($\lt$) for (a) screw and (b) edge dislocations in various bcc metals is shown as a function of $\bm=v\sqrt{\rho/\bar\m}$, and  compared to the isotropic limit (dashed gray curve).}
 \label{fig:linetension-bcc}
\end{figure}

The line tension of dislocations in bcc metals, on the other hand, exhibits a significantly more pronounced dependence on the elastic anisotropy, see figure~\ref{fig:linetension-bcc}.
Starting with figure~\ref{fig:linetension-bcc} (a),
nominally screw bcc dislocations, niobium and tungsten follow a similar trend to the isotropic curve, specifically, they exhibit only a single diverging term.
Both tantalum and more so iron contain at least two diverging terms of opposite sign.
Iron exhibits another peculiar feature in screw dislocations between $0.65\lessapprox\bm\lessapprox0.78$:
The small dip we saw in the dislocation line tension for fcc metals is much more pronounced in bcc iron and extends far into the negative region making straight screw dislocations unstable.
Next, turning our attention to figure~\ref{fig:linetension-bcc} (b),
the dashed isotropic curve suggests that with increasing velocity nominally edge dislocations become unstable.
Surprisingly however, this is only the case with tungsten ($A = 1.01$).
The other three (Fe, Nb, Ta) more anisotropic metals diverge positively, hence becoming infinitely stable at a scaled velocity which scales inversely with the degree of anisotropy;
For these three more anisotropic metals the critical velocity is $\bm \leq 0.9$.
With the exception of tungsten, figure~\ref{fig:linetension-bcc} suggests that all four bcc metals investigated exhibit diverging positive line tensions with increasing subsonic velocities for both nominally edge and screw dislocations.

\begin{figure}[h!t!b]
 \centering
 \includegraphics[width=\textwidth]{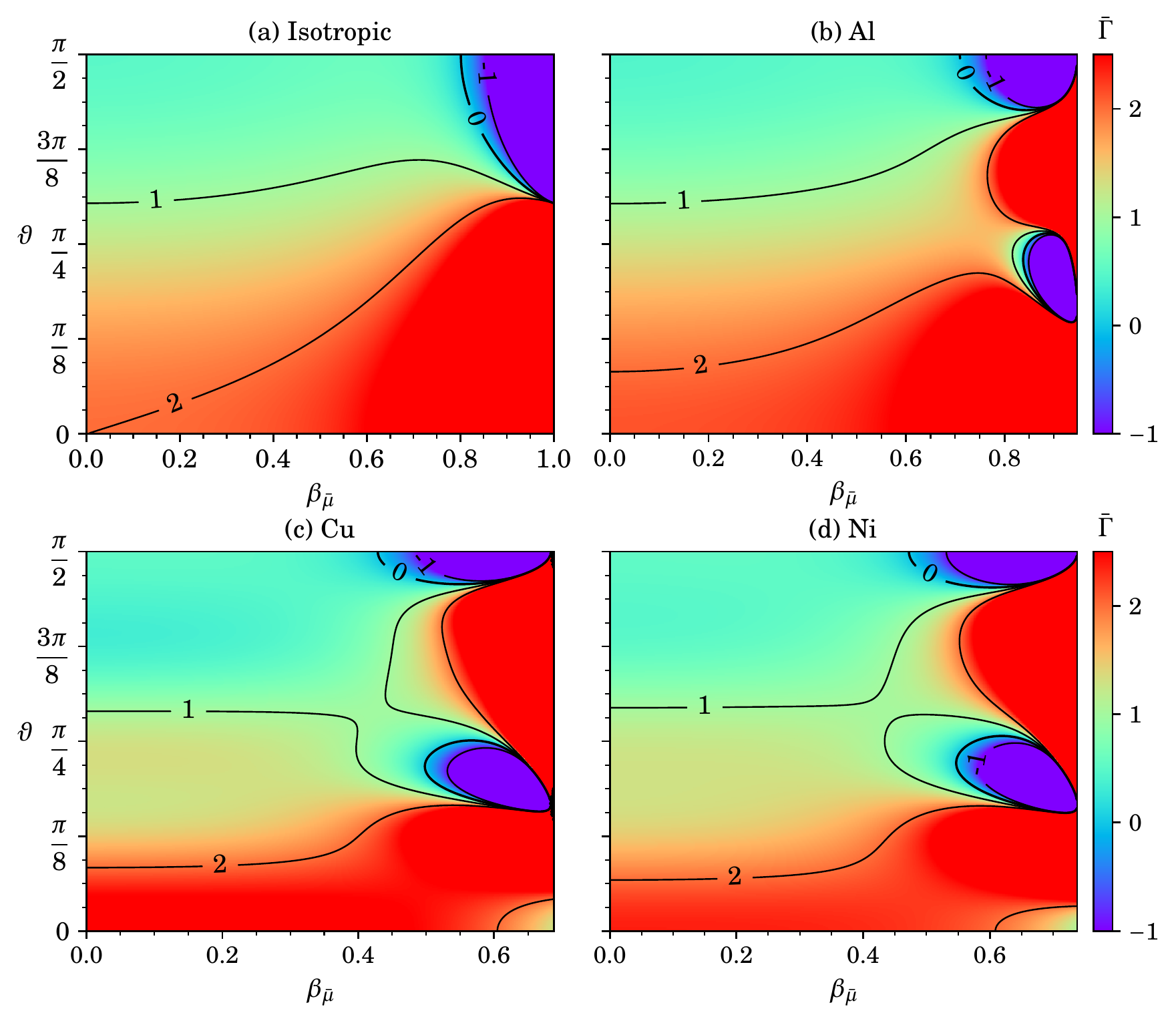}
 \caption{%
 The line tension prelogarithmic factor ($\lt$) for dislocations of arbitrary orientation is color-coded as a function of character angle ($\vth$) and scaled dislocation velocity ($\bm$) for the cases of (a) isotropic, (b) Al, (c) Cu, and (d) Ni.
 The angle $\vth=0$ corresponds to a pure screw dislocation and $\vth=\pm\pi/2$ corresponds to a pure edge dislocation.
 Regions of instability are indicated ($\lt<0$) in blue, and stable regimes ($\lt>0$) are indicated in red;
 For fcc and isotropic crystals, the dependence of $\lt$ on $\vth$ is symmetric about $\vth=0$, i.e. the negative $\vth$-range produces a mirror image of this plot.}
 \label{fig:linetension-fcc_contour}
\end{figure}

\begin{figure}[h!t!b]
 \centering
 \includegraphics[width=\textwidth]{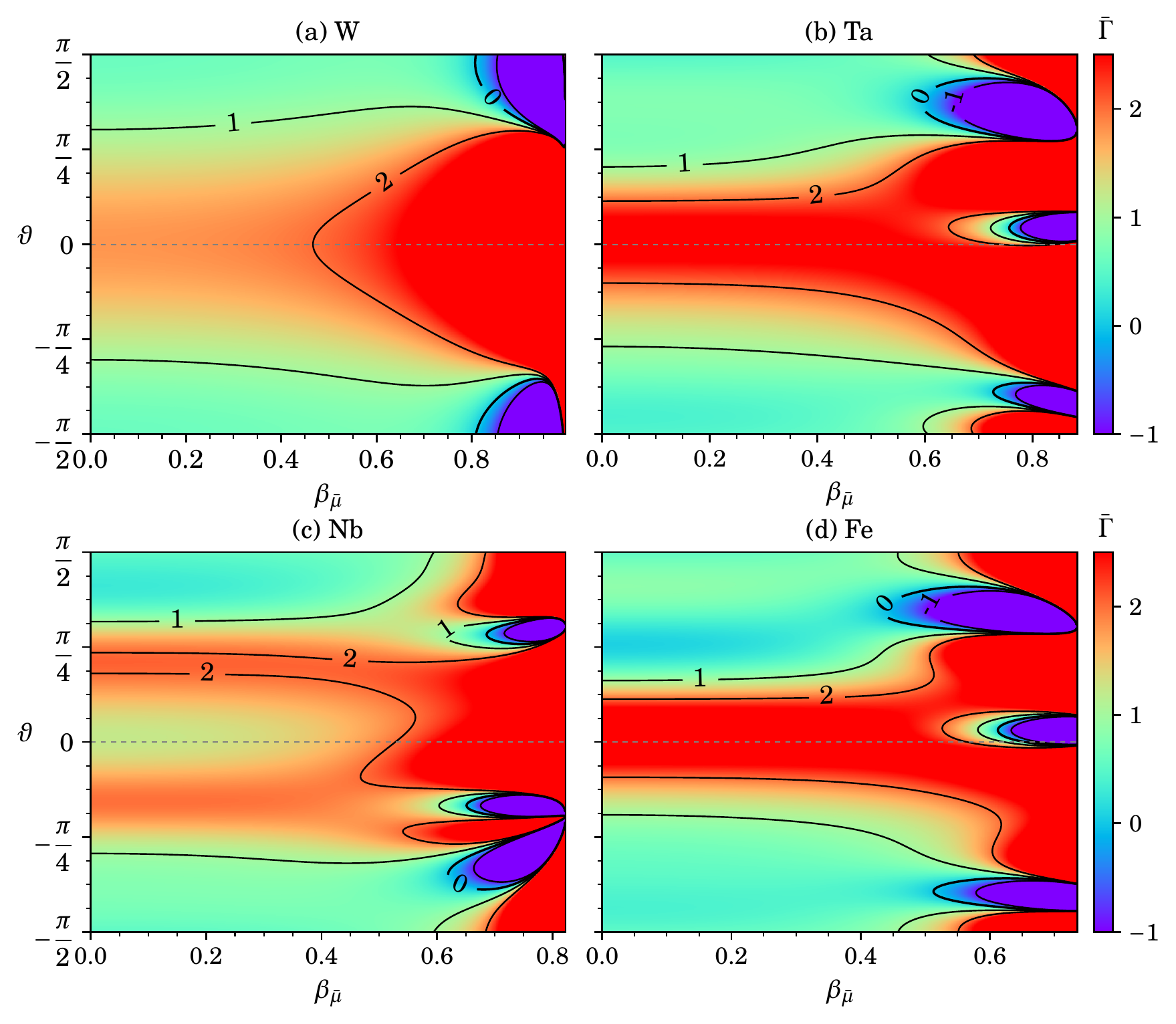}
 \caption{%
 The line tension prelogarithmic factor ($\lt$) for dislocations of arbitrary orientation is color-coded as a function of character angle ($\vth$) and scaled dislocation velocity ($\bm$) for the cases of (a) W, (b) Ta, (c) Nb, and (d) Fe.
 Regions of instability are indicated ($\lt<0$) in blue, and stable regimes ($\lt>0$) are indicated in red;
 For bcc crystals, the dependence of $\lt$ on $\vth$ is not symmetric about $\vth=0$, as seen within the figures (W, Ta, Nb, Fe).}
 \label{fig:linetension-bcc_contour}
\end{figure}

\subsection{Line tension dependence on velocity and character angle}

The figures for pure screw and edge dislocations (figures~\ref{fig:linetension-fcc} and \ref{fig:linetension-bcc}) are limited to the two extremes of the dislocation character in $\vth$ and do not capture potentially significant deviations from isotropy that may arise for mixed characters of dislocations.
To cover the entire gamut of parameter space for the line tension with respect to $\vth$ and $\bm$ for cubic anisotropic materials, we show the line tension for all seven metals investigated within this study
(i.e. figures~\ref{fig:linetension-fcc_contour} and \ref{fig:linetension-bcc_contour})
as a function of both the scaled velocity ($\bm$) as well as dislocation character ($\vth$).
Compared to elastic isotropy (i.e. figure~\ref{fig:linetension-fcc_contour} (a)), these figures exhibit a rich structure.

In organizing contour figures of $\lt$, in addition to $\bm$, we must consider the range of the character angle, $\vth$.
A physical interpretation of $\vth$ dictates that $\lt$ must be $\pi$-periodic in $\vth$, since $\lt$ should be insensitive to the direction of the line orientation ($\pm \mathbf{t}$).
Furthermore, within fcc crystals, the orientation of the Burgers vector ($\vec{b}^\txt{fcc} = (1,1,0)b/\sqrt{2}$) dictates that the dislocation line energy must be symmetric with respect to rotations of the dislocation line of $\pm \d\vth$ with respect to $\vec{b}^{\txt{fcc}}$ contained within the slip plane, $\vec{n}_0^\txt{fcc}$.
For bcc crystals, however, the dependence on $\vth$ is more subtle.
Physically, considering a small increment of $\pm \delta \vth$ of the dislocation line with respect to the bcc Burgers vector direction ($\vec{b}^\txt{bcc} = (1,-1,1)b/\sqrt{3}$), the orientation of the dislocation line is not symmetric with respect to its orientation within the bcc lattice, despite being symmetric with respect to the Burgers vector direction within the same bcc lattice.
This is because the slip plane ($\vec{n}_0^\txt{bcc}$) within a bcc unit lattice has an aspect ratio of $\sqrt{2}$.
We may succinctly express the lack of mirror symmetry in the line tension, which is \emph{only} present for the bcc materials investigated within this study, by expanding the zero velocity dislocation line energy about $\vth=0$, as
\begin{align}
\vec{b}^\txt{fcc}\cdot\mat{B}[\vec{t}^\txt{fcc}(\delta \vth)]\cdot\vec{b}^\txt{fcc} \approx \vec{b}^\txt{fcc}\cdot\mat{B}[\vec{t}^\txt{fcc}(0)]\cdot\vec{b}^\txt{fcc}+\sum_{n=0}^{\infty}\mathcal{O}(\delta \vth)^{2n}\,, \nn\\
\vec{b}^\txt{bcc}\cdot\mat{B}[\vec{t}^\txt{bcc}(\delta \vth)]\cdot\vec{b}^\txt{bcc} \approx \vec{b}^\txt{bcc}\cdot\mat{B}[\vec{t}^\txt{bcc}(0)]\cdot\vec{b}^\txt{bcc}+\sum_{n=0}^{\infty}\mathcal{O}(\delta \vth)^{n}\,
\end{align}
Since \emph{only} the fcc line energy prelogarithmic factor, $\vec{b}\cdot\mat{B}\cdot\vec{b}$ contains \emph{only} even terms in $\delta \vth$ of $\mathcal{O}(2n)$, we expect $\lt$ to be symmetric about $\vth=0$ for fcc materials \emph{only}, and we consider  $0 \leq \vth \leq \pi/2$.
With this symmetry lacking for the bcc materials which we examine, we extend the range from $-\pi/2 \leq \vth \leq \pi/2$.
We have numerically verified that these are the appropriate ranges.
In addition, we note that the zero velocity limit was previously computed for copper, nickel, niobium, and iron in ref.~\cite{Barnett:1972}, and our generalized results are in qualitative agreement with that earlier work at $\bm=0$.
Finally, in describing the contour plots for bcc metals, figures~\ref{fig:linetension-bcc_contour}, we note that our choice of slip system influences the dependence on $\vth$;
With an alternate slip plane, e.g.,
$\vec{b}^\txt{bcc}=\frac{b}{\sqrt{3}}\left(1,+1,1\right)$,
$\vec{n}_0^\txt{bcc}=\frac{1}{\sqrt{2}}\left(1,-1,0\right)$,
as opposed to \eqnref{eq:slipplanes-bcc}, the $\vth$ dependence within these plots would be reversed, i.e. $\vth\to-\vth$.

We proceed by investigating the line tension of several fcc metals, aluminum which is considered to be a ``fairly isotropic'' fcc metal exhibits a second, additional, unstable region at high velocities for $\vth$ close to $\pi/5$, see figure~\ref{fig:linetension-fcc_contour} (b).
With increasing anisotropy, the additional unstable region becomes even more pronounced, e.g., copper (figure~\ref{fig:linetension-fcc_contour} (c)) and nickel (figure~\ref{fig:linetension-fcc_contour} (d)).
All three figures~\ref{fig:linetension-fcc_contour} (b)--(d) are shown for $\bm\leq v_\txt{t}^{110}/v_{\bar\m}$, i.e. velocities below and including the lowest divergence which for fcc metals occurs for pure edge dislocations \emph{only}.
%%% numbers:
% vcrit(Al) = 0.948489
% vcrit(Cu) = 0.683821
% vcrit(Ni) = 0.738693
%%%%

%%%
Due to the differences in geometry, the line tension of mixed dislocations in bcc metals first diverges at the lowest shear wave speed at a character angle of $\vth=\arctan\left(\sqrt{2}\right)$.
All four figures~\ref{fig:linetension-bcc_contour} are therefore shown for $\bm$ below that speed, i.e. $\bm\leq \sqrt{(c_{11}-c_{12})/2\bar\m}$ for Fe, Ta, and W, and $\bm\leq \sqrt{c_{44}/\bar\m}$ for Nb.
%%% numbers:
% vcrit(Fe) = 0.735446
% vcrit(Nb) = 0.823501
% vcrit(Ta) = 0.883799
% vcrit(W) = 0.997541
%%%%%
%
We now turn our attention to those figures:
Even the almost isotropic tungsten (figure~\ref{fig:linetension-bcc_contour} (a)) exhibits some deviations in the stability of mixed ``almost edge'' dislocations in the high velocity regime compared to the truly isotropic case (figure~\ref{fig:linetension-fcc_contour} (a)).
Within tantalum, pure screw and edge dislocations appear stable, see figure~\ref{fig:linetension-bcc}.
However, upon inspection of figure \ref{fig:linetension-bcc_contour} (b), the existent unstable region splits into two parts which move away from angles characterizing pure edge, as seen within the figure.
Additionally, a third unstable region forms close to the screw configuration at angles $\vth>0$, and this explains the small decrease we see in the nominally screw case shown in figure~\ref{fig:linetension-bcc} (a).
In iron, which is even more anisotropic, the three unstable regions extend further into the lower velocity regime, and the near screw instability overlaps more with the pure screw configuration.
Again, these features were not visible in the figure~\ref{fig:linetension-bcc} for pure screw and edge.
What was visible already there, however, is the development of the third unstable region at small angles $\vth$ and moderately high velocities leading to the very pronounced decrease in the line tension for nominally screw dislocations.
As $\bm\to1$, those (almost) screw dislocations with small positive $\vth$ become stable once more.
Despite an unknown constant in the line tension which may shift the whole graph, it seems unlikely that this decrease (cf. figure~\ref{fig:linetension-bcc} (a)) will move into the positive region entirely, considering that for even small $\vth$ that dip extends orders of magnitude further into negative values.

Niobium, finally, behaves slightly different than the other bcc metals we have discussed thus far:
The two unstable regions in the range $-\pi/2<\vth<0$ have almost connected to one near the lowest transverse shear wave velocity $\b_{\bar \mu}\to\sqrt{c_{44}/\bar\mu}\approx0.82$ and are located at intermediate (negative) angles $\vth$, see figure~\ref{fig:linetension-bcc_contour} (c).
The reason this figure looks markedly different is that niobium has an anisotropic Zener ratio $A$ which is smaller than one (in contrast to all other metals we have discussed).
Once more, this feature was not visible in the plots for pure screw and edge dislocations (figure~\ref{fig:linetension-bcc}) which, without further inspection, may give the illusion of all Nb dislocations being stable, independent of subsonic velocity.

Lastly, we emphasize that because all elastic wave speeds in anisotropic crystals are direction dependent,
defining a border between ``subsonic'' and ``transonic'' is nebulous:
At sufficiently high velocities, the dislocation will be moving faster than the lowest shear wave speed with respect to one direction in the crystal, but not with respect to the lowest shear wave speed with respect to another direction.
We have also seen, that the divergence in the dislocation field is not necessarily equal to the lowest shear wave speed associated to the direction of dislocation motion.
Therefore, it makes more sense to distinguish between motions below and above the ``critical'' velocity where displacement gradients of dislocations are divergent, and this velocity depends upon the character angle $\vth$ of the dislocation.
Within this work, we hence use the term ``transonic'' as referring to a velocity above this critical velocity, a distinction which is unambiguously defined.
Since according to our table~\ref{tab:velocities} this velocity is very close to the lowest shear wave speed associated to the direction of dislocation motion, our slight abuse of the term ``transonic'' seems justified.

\subsection{Discussion and comparison to molecular dynamics simulations}

Comparing our results ($\lt(\vth,v)$) to earlier work, our figures~\ref{fig:linetension-fcc_contour} are in agreement with the notion of instabilities within MD simulations for pure edge dislocations \textit{only}, where ``jumps'' into the transonic regime were observed; Screw dislocations transitioned smoothly into the transonic regime)~\cite{Daphalapurkar:2014,Marian:2006,Oren:2017,Wang:2008,Gilbert:2011,Queyreau:2011}.
Within those works it was assumed that in analogy to the isotropic case divergences would occur at the lowest shear wave speed corresponding to the direction of motion of the dislocation.
As we have demonstrated, this is not necessarily the case for pure screw and edge dislocations and the actual critical velocity follows from \eqnref{eq:def-vcrit}, and therefore depends on the geometry of the crystal, the slip plane, and on the character of dislocation.
Typically, the critical velocity and the lowest shear wave speed corresponding to the direction of dislocation motion are very close together (see table~\ref{tab:velocities}), and are difficult to discern within the resolution of the MD simulations without detailed knowledge of the displacement gradient field as employed within this work.
What we presently have found for the line tension prelogarithmic factor ($\lt(\vth,v)$) of mixed dislocations, is that they diverge at the lowest shear wave speed only for a particular character angle, and for geometrical reasons, this angle is $\vth\to\pm\pi/2$ for fcc metals and $\vth\to\arctan\left(\sqrt{2}\right)$ for bcc metals.

In a further effort towards quantitative comparison with available data, we initially focus on recent MD simulations performed by Oren and coworkers~\cite{Oren:2017}.
The simulations were conducted on moving steady state nominally screw and edge dislocations within a copper sample at velocities both approaching, and exceeding the estimated first shear wave barrier.
As with the other simulations examined, the screw dislocations gradually transitioned from subsonic to transonic motion with increasing applied stress.
On the contrary, the nominally edge dislocation simulations exhibited the characteristic discontinuity in velocity versus stress at approximately the first shear wave barrier, which was recorded at a velocity of $v_\txt{crit} \approx 1.63$ km/s and is close to our computed value of $v_\txt{crit} \approx 1.61$ km/s (table~\ref{tab:velocities}).
The authors of~\cite{Oren:2017} graphically report the formation and expansion of nuclei along the dislocation line at this velocity as the mechanism for transonic transition (for dislocations longer than $\sim 10|\vec{b}|$).
We identify this mechanism as an instability, and proceed to apply our line tension analysis.
However, observing figure~\ref{fig:linetension-fcc} (b),
we notice that the line tension approaches zero at a velocity closer to 1 km/s.
We attribute this difference both to the existence of a constant ($\Gamma_0$) which does not scale logarithmically with the dislocation line length, and also the energetic interaction of the two separated partial dislocations~\cite{Pueschl:1987} which is beyond the scope of our analyses.
For example, a positive constant would translate the whole curve upwards and hence the line tension would become negative at a higher velocity which is closer to the reported critical velocity.

The same conclusion may be drawn for nickel when comparing our results to the MD simulations of~\cite{Daphalapurkar:2014,Marian:2006}.
Also within these MD simulations a discontinuity in edge dislocation velocity versus stress is observed closer to the critical velocity than within our figure~\ref{fig:linetension-fcc} (b).
Similarly, due to the omission of an unknown constant in our line tension calculation, a comparison of results can only be made qualitatively.
Even a moderate shift of our curves by a constant can move the crossing of the zero axis significantly closer to the critical velocity, especially since the asymptotes become fairly steep.

In ref.~\cite{Olmsted:2005}, MD simulations for Al and Ni lead to the conclusion that the discontinuity in velocity happens slightly below the lowest shear wave speed (which deviates from ours because the interatomic potential employed by the authors yields differing values for their elastic constants) corresponding to the direction of dislocation motion, and therefore agrees somewhat better with our present study.

\section{Conclusion and Outlook}
\label{sec:conclusion}
%%%%%%%%%%%%%%%%%%%%%%%%

We have presented a generalized analysis of the line tension prelogarithmic factor ($\lt(\vth,v)$) of dislocations~\cite{Bacon:1980} within anisotropic media for both an arbitrary character angle, within the slip plane, as well as a constant, finite velocity.
We have applied our analyses towards computing the line tension of a suite of fcc and bcc structured crystals over a range of character angles and subsonic velocities.
Upon including full elastic anisotropy, we find that certain character angles exhibit a diverging, negative line tension with increasing velocity, suggesting that the concept of straight dislocations within these regimes of $\vth$-$v$ space is energetically unfavorable, and likely should not be applied.
In addition, we find that the lowest material shear wave velocity which is responsible for the diverging line tension is not necessarily parallel to the direction of dislocation motion, as often assumed~\cite{Olmsted:2005,Oren:2017,Daphalapurkar:2014,Marian:2006}, but rather contained within the plane defined by the line direction of the dislocation under consideration.
Finally, we have demonstrated that the line tension of bcc structured materials lacks mirror symmetry (independent of velocity) with respect to the dislocation character angle about the nominally screw, or edge orientation.
We attribute this asymmetry in the line tension to the geometry of the bcc crystalline lattice (both Burgers vector and 
slip plane orientation) which lacks symmetry with respect to the reference cubic crystal axes.

Based upon our bcc results, figures~\ref{fig:linetension-bcc_contour} (b)--(d), we may make a prediction about bcc structured Fe, Nb, and Ta.
We expect that MD simulations will show that pure edge dislocations in Fe, Nb, and Ta do not ``jump'' into the transonic regime with increasing velocity;
However, mixed dislocations within a range of angles $\vth$ do;
This is particular to the bcc structure, and dependent upon the inclusion of elastic anisotropy (cf. figure \ref{fig:linetension-bcc} (b)).
We interpret ``transonic'' as velocities above those at which the displacement gradients of steady-state moving dislocations diverge.
We emphasize that this speed, with the inclusion of elastic anisotropy, is typically close to, but less than the lowest shear wave speed corresponding to the direction of dislocation movement.
The two speeds coincide only for select character angles ($\vth$) in fcc and bcc metals.

Both the analyses and the results of this work delineate regimes of forbidden velocities within a two dimensional dislocation framework, where dislocations are interpreted to be straight, infinite entities.
We anticipate that both the existence and identification of these regimes will aid in enhancing our understanding of dynamic plasticity within fully anisotropic crystals under high rates of loading.

\subsection*{Acknowledgements}
%%%%%%%%%%%%%%%%%%%%%%%%%%%%%%%%%%%%%%%%%%

We thank D. J. Luscher for enlightening discussions.
B. A. S. thanks Joshua Crone for fruitful discussions.
We also thank the anonymous referee for valuable comments.
This work was performed in part under the auspices of the U.S. Department of Energy under contract DE-AC52-06NA25396.
In particular, the authors are grateful for the support of the Advanced Simulation and Computing, Physics and Engineering Models Program.
In addition, research was sponsored by the Army Research Laboratory and was accomplished under Cooperative Agreement Number W911NF-17-2-0224.
The views and conclusions contained in this document are those of the authors and should not be interpreted as representing the official policies, either expressed or implied, of the Army Research Laboratory or the U.S. Government.
The U.S. Government is authorized to reproduce and distribute reprints for Government purposes notwithstanding any copyright notation herein.

%%%%%%%%%%%%%%%%%%%%%%%%%%%%%%%%%%%%%%%%%%%
\bibliographystyle{utphys-custom}
\bibliography{linetension}
%%%%%%%%%%%%%%%%%%%%%%%%%%%%%%%%%%%%%%%%%%%

\end{document}